\documentclass[prd,onecolumn,notitlepage,eqsecnum,nofootinbib,floatfix]{revtex4-1}
\usepackage{amssymb}
\usepackage{amsmath}
\usepackage{amsfonts} 
\usepackage{bm}
\usepackage{graphicx}
\DeclareSymbolFont{EulerScript}{U}{eus}{m}{n}
\SetSymbolFont{EulerScript}{bold}{U}{eus}{b}{n}
\DeclareSymbolFontAlphabet\scrpt{EulerScript}
\allowdisplaybreaks
\newcommand{\arccot}{{\mbox{arccot}}\,} 
\newcommand{\Lie}{{\pounds}} 
\newcommand{\KK}{{\scrpt K}} 
\newcommand{\F}{{\scrpt F}} 
\newcommand{\R}{{\scrpt R}} 
\newcommand{\SSS}{{\scrpt S}} 
\newcommand{\bkt}[2]{{\langle #1 | #2 \rangle}} 

\newcommand{\gothp}{\mathfrak{p}} 
\newcommand{\gothq}{\mathfrak{q}} 
\newcommand{\gotho}{\mathfrak{o}} 
\newcommand{\gothh}{\mathfrak{h}} 
\newcommand{\gothN}{\mathfrak{N}} 
\newcommand{\e}[1]{{\times 10^{#1}}}   
\begin{document}
\title{Tidal driving of inertial modes of Maclaurin spheroids}  
\author{Eric Poisson and Cyann Buisson}  
\affiliation{Department of Physics, University of Guelph, Guelph,
  Ontario, N1G 2W1, Canada} 
\date{October 7, 2020} 
\begin{abstract} 
We examine the inertial modes of vibration of a Maclaurin spheroid and determine how they are driven by an external tidal field, either Newtonian (gravitoelectric) or post-Newtonian (gravitomagnetic). The context and motivation for this work come from the realization that inertial modes of rotating neutron stars can be resonantly excited in binary inspirals, with a measurable effect on the phasing of the emitted gravitational waves. We aim to provide additional insights into this phenomenon. We calculate how the frequencies of the relevant inertial modes, and their overlap integrals with the tidal forces, vary as the star's rotation rate increases, spanning almost the full range between zero rotation and the mass-shedding limit. We prove that a single inertial mode couples to a Newtonian tidal field; overlap integrals with all other modes vanish. We prove also that four inertial modes couple to a post-Newtonian, gravitomagnetic tidal field; overlap integrals with all other modes vanish. Finally, we determine the rather extreme conditions under which the gravitoelectric driving of inertial modes dominates over the gravitomagnetic driving.  
\end{abstract} 
\maketitle

\section{Introduction and summary} 
\label{sec:intro} 

\subsection{Resonant driving of inertial modes in neutron-star binary inspirals} 

The tidal interaction between a neutron star and a companion body in a coalescing compact binary can bring the star's normal modes of vibration into resonance, which leads to a significant exchange of energy between the star and the orbital motion. This resonant interaction alters the orbit and leaves an imprint on the phasing of the emitted gravitational waves. In principle the phase shift can be measured by interferometric detectors, with the promise to reveal details of the neutron-star interior and the nature of nuclear matter at high densities. 

A number of modes can be implicated in the tidal interaction. Early studies \cite{lai:94, reisenegger-goldreich:94, shibata:94, kokkotas-schafer:95, lai:97, ho-lai:99} focused on $f$-modes and $g$-modes, with the conclusion that $f$-modes cannot become resonant prior to merger (because their frequencies are too high, beyond the LIGO/Virgo frequency band), and that while $g$-modes can be resonantly excited, their coupling to the tidal force is too weak to lead to measurable effects on the orbit. Another possibility, identified recently by Pan {\it et al.}\ \cite{pan-etal:20}, are interface modes \cite{mcdermott-vanhorn-hansen:88} that result from an interaction between the fluid core and solid crust of a neutron star. While these modes do not typically achieve resonance, they nevertheless produce noticeable phasing effects in the gravitational waves; these could be revealed in future improvements of the LIGO/Virgo detectors. 

Another set of modes that can participate in a tidal interaction is the class of {\it inertial modes} of a rotating neutron star. These modes, which were first identified by Lockitch and Friedman \cite{lockitch-friedman:99}, are predominantly perturbations of the star's velocity field, subjected to a restoring force supplied by the Coriolis force. These modes are especially promising for a resonant interaction, because their frequencies are proportional to the star's rotational angular velocity; they lie within the LIGO/Virgo frequency band when the star spins at a rate comparable to $100\ \mbox{Hz}$. The inertial modes include $r$-modes \cite{papaloizou-pringle:78, provost-berthomieu-rocca:81, saio:82, smeyers-martens:83} as a purely axial subclass. 

The interaction between inertial modes and a Newtonian tidal field was first explored by Alexander \cite{alexander:87, alexander:88}. Lai and Wu \cite{lai-wu:06} (see also Ref.~\cite{xu-lai:17}) concluded that a noticeable effect on the gravitational waves would require rotational frequencies well beyond $100\ \mbox{Hz}$. Such frequencies may seem unlikely 
given the range of known pulsar frequencies, but they cannot be excluded; a 316 Hz binary pulsar (PSR J1717+4308a) was recently discovered \cite{pan-ransom-etal:20}, and the fastest known isolated pulsar (PSR J1748-2446ad) spins at a rate of 716 Hz \cite{hessels:06}. Flanagan and Racine \cite{flanagan-racine:07}, however, pointed out that inertial modes interact more strongly with a post-Newtonian tidal field, in spite of the inherent suppression by a factor of $vv'/c^2$ relative to the Newtonian field; here $v$ is the rotational velocity, $v'$ the orbital velocity, and $c$ the speed of light. By virtue of this stronger coupling, Flanagan and Racine predicted phase shifts that could be within reach in near-future improvements of the LIGO/Virgo detectors. 

The tidal driving of inertial modes in a binary inspiral was recently revisited by Poisson in Ref.~\cite{poisson:20a}. While Flanagan and Racine focused their attention on a single $r$-mode, Poisson showed that a fuller set of four inertial modes (including the $r$-mode) participates in the tidal interaction, each mode producing a distinct resonance. The total gravitational phase shift accumulated during these resonances can be expected to be in an interval between $0.02$ and $0.04$ radians when the neutron star's spin is closely aligned with the orbital angular momentum, or in an interval between $0.08$ and $0.16$ radians when the spin is nearly anti-aligned. These estimates should be multiplied by two when the companion is also a rotating neutron star with a comparable spin. Though small, such phase shifts should become measurable within the current decade, thanks to incremental improvements of the LIGO/Virgo instruments. In the longer term, the resonant tidal driving of inertial modes will have to be incorporated in waveform models for an accurate measurement of the waves by a third generation of detectors (Cosmic Explorer \cite{CosmicExplorer}; Einstein Telescope \cite{EinsteinTelescope}). 

Our purpose with this paper is to provide additional insights into the tidal driving of inertial modes of rotating neutron stars. We explore two themes. In the first, we determine how much the mode frequencies and overlap integrals with a tidal force vary with $\Omega$, the star's rotational velocity. The description of inertial modes supplied by Lockitch and Friedman \cite{lockitch-friedman:99} is limited by a small-$\Omega$ approximation; here we calculate the frequencies and overlap integrals to all orders in $\Omega$. In the second theme, we examine the suppression of the Newtonian tidal driving relative to the post-Newtonian driving, and determine the (rather extreme) circumstances under which Newtonian tides dominate. Both themes require us to go beyond the small-$\Omega$ approximation in the description of inertial modes. For this we rely on the beautiful analysis of Lindblom and Ipser \cite{lindblom-ipser:99}, who constructed (following Bryan \cite{bryan:89}) the complete class of inertial modes for Maclaurin spheroids. The price to pay for going beyond the small-$\Omega$ approximation is therefore to adopt a rather unrealistic stellar model, one with a constant mass density. The benefit is that the inertial modes can be determined exactly to all orders in $\Omega$. We expect that our results, based on the Maclaurin spheroid, give reliable order-of-magnitude estimates for more realistic stellar models.    

The driving of inertial modes of Maclaurin spheroids by a Newtonian tidal field was studied previously by Braviner and Ogilvie \cite{braviner-ogilvie:14, braviner-ogilvie:15}, in the context of the tidal dynamics of a star-planet system; there is minimal overlap between their work and ours. The role of inertial modes in the tidal heating of Jupiter was examined by Wu \cite{wu:05a, wu:05b}. The potential impact of inertial modes on electromagnetic phenomena (such as precursor flares in gamma-ray bursts) was recently explored by Suvorov and Kokkotas \cite{suvorov-kokkotas:20}, following up on previous work by Tsang {\it et al.}\ \cite{tsang-etal:12}. 

\subsection{First theme: Beyond the small-$\Omega$ approximation} 

The small-$\Omega$ approximation is an expansion in powers of 
\begin{equation} 
\epsilon := \frac{\Omega}{\sqrt{\pi G \rho}} 
= 5.32 \times 10^{-2} \biggl( \frac{1.4\ M_\odot}{M} \biggr)
\biggl( \frac{R}{10\ \mbox{km}} \biggr)^{3/2} 
\biggl( \frac{f_{\rm rot}}{100\ \mbox{Hz}} \biggr), 
\label{epsilon_def} 
\end{equation} 
where $\Omega$ is the rotational angular velocity of the neutron star, $\rho = 3M/(4\pi R^3)$ its averaged mass density (with $M$ the mass and $R$ the radius), and $G$ is the gravitational constant; in the second expression we introduced $f_{\rm rot} := \Omega/(2\pi)$ as the rotational frequency and inserted fiducial values for the mass, radius, and frequency of a neutron star. The ratio $\epsilon$ approaches unity when $\Omega$ becomes comparable to the mass-shedding limit of $\sqrt{\pi G\rho}$, at which an element of mass on the surface rotates at approximately the Keplerian frequency. The expression of Eq.~(\ref{epsilon_def}) reveals that $\epsilon$ is small for the neutron stars considered here, which spin at a rate comparable to $100\ \mbox{Hz}$. For a star with $f_{\rm rot} \simeq 1.25\ \mbox{kHz}$, however, $\epsilon$ is approximately equal to $0.664$, which marks a bifurcation on the sequence of Maclaurin spheroids, where it connects with a sequence of Jacobi ellipsoids (see Chapter 5 of Ref.~\cite{chandrasekhar:69} or Sec.~2.3.2 of Ref.~\cite{poisson-will:14}); this also marks the onset of a secular instability. In the following we shall take $\epsilon \simeq 0.67$ as an upper bound on $\Omega/\sqrt{\pi G \rho}$. 

In the small-$\Omega$ approximation of Lockitch and Friedman \cite{lockitch-friedman:99}, the description of inertial modes comes with a perturbation in the star's velocity field of order unity (with corrections of order $\epsilon^2$), while the perturbations in pressure, density, and gravitational potential scale as $\epsilon^2$; the mode eigenfrequencies are directly proportional to $\Omega$ (also with corrections of order $\epsilon^2$). Overlap integrals between the four dominant inertial modes and a post-Newtonian, gravitomagnetic tidal force produced by an orbiting companion were computed in Ref.~\cite{poisson:20a}, also in the context of the small-$\Omega$ approximation. The dominant modes consist of one mode with $m = 0$ (actually a pair of complex-conjugate modes), two modes with $m = 1$ (one with positive frequency, the other with negative frequency), and an $r$-mode with $m = 2$; the integer $m$ characterizes the $e^{im\phi}$ behavior of the perturbation, with $\phi$ denoting the azimuthal angle around the axis of rotation.  

\begin{table} 
\caption{\label{tab:range} Range of variation of mode frequencies and overlap integrals with a post-Newtonian, gravitomagnetic tidal force.} 
\begin{ruledtabular} 
\begin{tabular}{lcc}
Mode & Frequency & Overlap integral \\ 
\hline 
$m = 0$ & 33\% & 52\% \\ 
$m = 1$ (positive frequency) & 61\% & 6\% \\ 
$m = 1$ (negative frequency) & 20\% & 100\% \\ 
$m = 2$ & 57\% & 27\% 
\end{tabular} 
\end{ruledtabular} 
\end{table} 

Our main goal with the first theme is to determine how the mode frequencies and overlap integrals vary when $\epsilon$ is no longer restricted to small values, but is allowed to span the interval $0 < \epsilon < 0.67$. We carry this out in 
Sec.~\ref{sec:results}, and our results are summarized in Table~\ref{tab:range}. Our main observation is that the frequencies and overlap integrals change appreciably; they either increase or decrease with increasing $\epsilon$. While the changes are slow so long as $\epsilon$ is small (because the deviations from leading order scale as $\epsilon^2$), they become rapid as $\epsilon$ increases beyond approximately $0.3$; with the fiducial values of Eq.~(\ref{epsilon_def}), this corresponds to $f_{\rm rot} > 560\ \mbox{Hz}$. For the expected range of rotational frequencies of binary neutron stars, the mode frequencies and overlap integrals do not vary much. 

In the context of the small-$\Omega$ approximation, Poisson \cite{poisson:20a} showed that for stellar models with a varying mass density, only four inertial modes produce appreciable overlap integrals with the post-Newtonian tidal force; the remaining modes give rise to integrals that are smaller by orders of magnitude. We show in Sec.~\ref{sec:vanishing} that for a Maclaurin spheroid, {\it all overlap integrals between inertial modes and a post-Newtonian, gravitomagnetic tidal force vanish, except for precisely four modes}. The statement is true for any value of $\Omega$, and is valid at leading quadrupole order in a multipole expansion of the tidal field. The exceptional modes are those designated below by $\ell = 3$ with $m = \{0, 1, 2\}$; there are two modes with $m=1$. In the small-$\Omega$ regime, the exceptional modes reduce to the dominant Lockitch-Friedman modes constructed for a star with constant density. 

\subsection{Second theme: Newtonian tidal driving of inertial modes} 

The post-Newtonian, gravitomagnetic tidal force created by a companion body is smaller than the Newtonian (gravitoelectric) force, but it typically produces a stronger driving of inertial modes. Our goal with the second theme is to examine this unexpected effect, and to identify the conditions under which the Newtonian driving would dominate. 

The Newtonian tidal force density results from a scalar potential $U$, and the overlap integral with a mode described by a Lagrangian displacement vector $\chi^a$ is 
\begin{equation} 
f^{\rm el} = \int \rho\, \chi^a \nabla_a U\, d{\cal V}, 
\end{equation} 
where $d{\cal V}$ is the element of volume, and the domain of integration is the region occupied by the unperturbed star. 
With $\chi^a \sim R$ (by an appropriate choice of normalization) and $U \sim GM' R^2/r^3$, where $M'$ is the mass of the companion body and $r$ the orbital radius, we obtain the naive estimate $f^{\rm el} \sim GMM' R^2/r^3$ for the overlap integral. To see why this is far off the mark, we integrate by parts to produce the alternative expression
\begin{equation} 
f^{\rm el} = \oint \rho\, U n_a \chi^a\, dS 
- \int U \nabla_a (\rho\, \chi^a)\, d{\cal V} 
\end{equation} 
for the overlap integral; the first integral is over the surface of the star, with $n_a$ the unit normal vector and $dS$ the element of surface area. Because the inertial modes satisfy $\nabla_a (\rho \chi^a) = 0$ (in the interior) and $n_a \chi^a = 0$ (on the surface) in the small-$\Omega$ approximation, we have that the overlap integral vanishes. This is the underlying reason for the weakness of the coupling between inertial modes and Newtonian tidal forces. 

To produce a more reliable estimate for the overlap integral, we must multiply our preceding one by a factor of $\epsilon^2$. This yields
\begin{equation} 
f^{\rm el} \sim \frac{M' \Omega^2 R^5}{r^3}. 
\end{equation} 
On the other hand, the post-Newtonian, gravitomagnetic tidal force acting on the rotating star is created by a vector potential $\bm{U}$ associated with the companion's orbital motion, and is closely analogous to the $c^{-2} \bm{v} \times \bm{B}$ force of electromagnetism, with $\bm{v}$ the star's rotational velocity and $\bm{B} = \bm{\nabla} \times \bm{U}$ the gravitomagnetic field. With $M' v'$ playing the role of mass current, we have that  
$\bm{U} \sim GM' v' R^2/r^3$, and we arrive at the estimate 
\begin{equation} 
f^{\rm mag} \sim \frac{G M M' v' \Omega R^3}{c^2 r^3} 
\end{equation} 
for the overlap integral. The ratio of gravitoelectric to gravitomagnetic tidal driving is therefore measured by 
\begin{equation} 
\R := \frac{f^{\rm el}}{f^{\rm mag}} \sim \frac{c^2 \Omega R^2}{GM v'} 
= \frac{f_{\rm rot}}{f_{\rm orb}} \frac{R}{r} \frac{R}{GM/c^2}, 
\label{ratio_estimate} 
\end{equation} 
where $f_{\rm orb}$ is the companion's orbital frequency. The estimate reveals that $\R$ is essentially the product of the ratio of rotational to orbital frequencies, the ratio of stellar to orbital radius, and the ratio of stellar to gravitational radius. In the context of a resonant tidal interaction in a compact binary, we have that $f_{\rm rot}$ is comparable to $f_{\rm orb}$, $R$ is much smaller than $r$, and $R$ is larger than $GM/c^2$. The actual value of $\R$ depends on the competition between $R/r$ and $R/(GM/c^2)$, but it also depends on the numerical factors that show up in an actual calculation of the overlap integrals. 

In Sec.~\ref{sec:ratio} we perform a detailed calculation of $\R$. The computation incorporates the numerical factors mentioned previously, but it also accounts for the number of modes implicated in the overlap integrals (a single mode for $f^{\rm el}$, four modes for $f^{\rm mag}$), and for the number of frequency components of the tidal fields. We obtain 
\begin{equation} 
\R = \frac{5}{2} \biggl( 1 + \frac{3}{2} \sqrt{3} \biggr)  \frac{1}{\Gamma} 
\frac{c^2 \Omega R^2}{GM v'} 
= (2\pi)^{3/2} \frac{5}{2} \biggl( 1 + \frac{3}{2} \sqrt{3} \biggr) \frac{1}{\Gamma}  
\frac{c^2 R^2}{(GM) (GM_{\rm tot})^{1/3}} \frac{f_{\rm rot}}{f_{\rm orb}^{1/3}}, 
\label{ratio_intro} 
\end{equation} 
where the numerical factors including $\Gamma^{-1}$ --- a function of $f_{\rm rot}$ and $f_{\rm orb}$ formally introduced in Eq.~(\ref{Gamma_def}) --- come from the precise calculation; in the second expression we related the orbital velocity $v'$ to the orbital frequency $f_{\rm orb}$, with the help of the binary's total mass $M_{\rm tot} = M + M'$. Inserting fiducial values for the radius, masses, and frequencies, this is 
\begin{equation} 
\R = \biggl( \frac{R}{10\ \mbox{km}} \biggr)^2 \biggl( \frac{1.4\ M_\odot}{M} \biggr) 
\biggl( \frac{2.8\ M_\odot}{M_{\rm tot}} \biggr)^{1/3}\, \SSS 
\end{equation} 
with 
\begin{equation} 
\SSS := 0.178 \biggl( \frac{25}{\Gamma} \biggr) \biggl( \frac{f_{\rm rot}}{100\ \mbox{Hz}} \biggr) 
\biggl( \frac{100\ \mbox{Hz}}{f_{\rm orb}} \biggr)^{1/3}. 
\label{S_intro} 
\end{equation} 
In Eq.~(\ref{S_intro}), the main dependence on the orbital and rotational frequencies comes from the overall factor of $f_{\rm rot}\, f_{\rm orb}^{-1/3}$. But the prefactor $\Gamma^{-1}$ contributes a significant modulation, as it varies by more than a factor of two over the considered range of frequencies; a typical value is $\Gamma \sim 25$.   

\begin{figure} 
\includegraphics[width=0.4\linewidth]{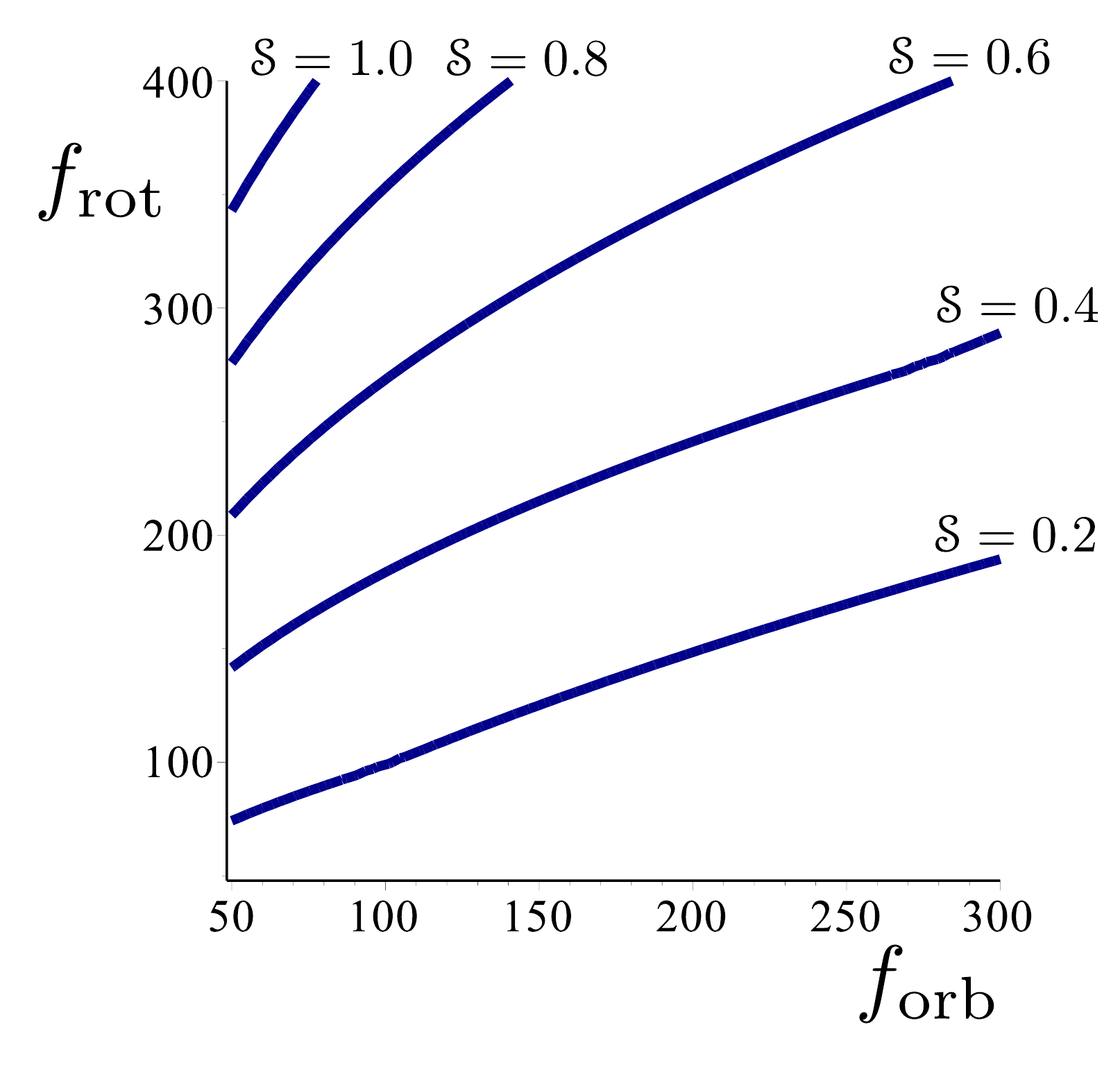}
\caption{Reduced ratio $\SSS$ of gravitoelectric to gravitomagnetic tidal driving. In ascending order from the lowest to highest curve, we have contours of $\SSS = \{ 0.2, 0.4, 0.6, 0.8, 1.0 \}$, respectively.} 
\label{fig:Ratio} 
\end{figure} 

In Fig.~\ref{fig:Ratio} we display a contour plot of the reduced ratio $\SSS$ of Eq.~(\ref{S_intro}). The salient point, our main conclusion within the second theme, is that $\SSS > 1$ requires a rotational frequency larger than approximately $350\ \mbox{Hz}$ when $f_{\rm orb} \simeq 50\ \mbox{Hz}$; and this threshold value increases rapidly (roughly like $f_{\rm rot} \propto f_{\rm orb}^3$) with increasing orbital frequency. For the rotational frequencies expected of binary inspirals, gravitomagnetic tidal driving of inertial modes will always dominate over gravitoelectric tidal driving. 

We mentioned in passing that $\R$ implicates a single inertial mode for the gravitoelectric tidal driving, against four for the gravitomagnetic driving. This is another key result within our second theme: we show in Sec.~\ref{sec:vanishing} that for a Maclaurin spheroid, {\it all overlap integrals between inertial modes and a Newtonian tidal force vanish, except for one single mode}. The statement is true for any value of $\Omega$, and is valid at leading quadrupole order in a multipole expansion of the tidal field. The exceptional mode is the $r$-mode designated below by $\ell = 2$ and $m = 1$. This mode has the same frequency for any value of $\Omega$; it is equal to $-\Omega$ in the star's corotating frame, and it vanishes in the nonrotating frame. 

\subsection{Organization of the paper}  

We begin in Sec.~\ref{sec:background} with a summary of essential background material from the literature on Maclaurin spheroids and their inertial modes. In Sec.~\ref{sec:driving} we introduce the standard formalism to describe the tidal driving of inertial modes, measured in terms of overlap integrals between the modes and the tidal force. We calculate all nonvanishing overlap integrals in Sec.~\ref{sec:results}, and in Sec.~\ref{sec:vanishing} we prove that all remaining integrals vanish. These two sections are devoted to an exploration of the first theme. We turn to the second theme in Sec.~\ref{sec:ratio}, where we calculate the ratio $\R$ of gravitoelectric to gravitomagnetic tidal driving. 

The paper features a large number of symbols that occur frequently; we provide a list in Appendix~\ref{sec:symbols}. In Appendix~\ref{sec:axi_zerof} we offer a thorough description of the axisymmetric, zero-frequency modes of a Maclaurin spheroid, and show that these do not couple to a gravitoelectric or gravitomagnetic tidal force. In Appendix~\ref{sec:legendre} we list a number of integrals involving Legendre polynomials and associated Legendre functions; these are required in the manipulations of Sec.~\ref{sec:vanishing}. 

The technical developments of this paper rely on two key sources: the construction of inertial modes of Maclaurin spheroids by Lindblom and Ipser \cite{lindblom-ipser:99}, which is summarized in Sec.~\ref{sec:background}, and the formalism of driven perturbations of rotating stars by Schenk {\it et al.}\ \cite{schenk-etal:01}, which is outlined in Sec.~\ref{sec:driving}. Poisson's recent work on gravitomagnetic tidal resonances in binary inspirals \cite{poisson:20a} provides additional context and motivation for the work carried out here. It also contains a more complete review of the relevant literature, as well as a long list of simplifying assumptions and a discussion of their limitations.  

\section{Background material} 
\label{sec:background} 

We collect some background material from the literature, with the intention to make this paper reasonably self-contained. Most of this is an executive summary of the 1999 paper by Lindblom and Ipser \cite{lindblom-ipser:99}.   

\subsection{Maclaurin spheroids} 
\label{sec:maclaurin} 

A Maclaurin spheroid is an axially symmetric body of uniform density that rotates rigidly with a constant angular velocity. The construction is detailed in Chapter 5 of Chandrasekhar's {\it Ellipsoidal Figures of Equilibrium} \cite{chandrasekhar:69}, or in Sec.~2.3.2 of Poisson and Will's {\it Gravity} \cite{poisson-will:14}; we summarize the main results here. We adopt the notation of Lindblom and Ipser \cite{lindblom-ipser:99}.  

The spheroid is characterized by a density $\rho$, a mass $M$, an angular velocity $\Omega$, a focal radius $a$, and an ellipticity parameter $\zeta_0$ that lies between $0$ and $\infty$; the nonrotating limit corresponds to $\zeta_0 \to \infty$. The angular velocity is related to the ellipticity parameter by 
\begin{equation} 
\Omega^2 = 2 \pi G \rho\, \zeta_0 \bigl[ (1 + 3\zeta_0^2) \arccot\zeta_0 - 3 \zeta_0 \bigr]. 
\label{O2}
\end{equation} 
The specific enthalpy $h := p/\rho$ is given by 
\begin{equation} 
h = h_c \biggl( 1 - \frac{x^2 + y^2}{R_e^2} - \frac{z^2}{R_p^2} \biggr), 
\label{enthalpy} 
\end{equation} 
where 
\begin{equation} 
h_c = 2\pi G \rho a^2 \zeta_0^2 (1 + \zeta_0^2)(1 - \zeta_0\, \arccot\zeta_0) 
\label{hc} 
\end{equation} 
is the central value, while 
\begin{equation} 
R_e := (1 + \zeta_0^2)^{1/2}\, a, \qquad 
R_p := \zeta_0\, a 
\label{Re_Rp}
\end{equation} 
are respectively the spheroid's equatorial and polar radii; $a$ acquires its meaning through these relations. The rotation axis is aligned with the $z$-axis, and the equator corresponds to the $x$-$y$ plane. The surface of the spheroid is defined to be the place where $h$ vanishes. 

The spheroid's mean radius $R$ is defined by $R^3 := R_e^2 R_p$ and is given by 
\begin{equation} 
R = \zeta_0^{1/3} (1 + \zeta_0^2)^{1/3}\, a. 
\label{mean_radius} 
\end{equation} 
The spheroid's volume is ${\cal V} = (4\pi/3) R^3$, and its mass is $M = \rho {\cal V}$. On a sequence of spheroids of constant density and mass but varying $\Omega$ (and therefore $\zeta_0$), we have that $R$ is constant and $a$ varies according to Eq.~(\ref{mean_radius}).  

The Maclaurin spheroid becomes secularly unstable (to the formation of a Jacobi ellipsoid) when $\zeta_0$ is smaller than the critical value of $0.480116$. This corresponds to an angular velocity $\Omega \simeq 0.664287\, \sqrt{\pi G\rho}$. 

In some calculations it is convenient to write 
\begin{equation} 
\rho = \rho_0 \Theta(h), 
\label{EOS} 
\end{equation} 
where $\rho_0$ is the constant density inside the spheroid, and $\Theta(h)$ is the Heaviside step function. We may interpret Eq.~(\ref{EOS}) as an equation of state $\rho = \rho(h)$. In this description, $h$ is positive inside the spheroid, and is taken to extend to negative values beyond the surface. In most equations it will not be necessary to distinguish between $\rho$ and $\rho_0$. But in equations involving the derivative of the density, Eq.~(\ref{EOS}) gives rise to terms proportional to a Dirac distribution; these play an important role in the determination of boundary conditions on the surface. 

\subsection{Coordinate systems} 
\label{sec:coordinates} 

We introduce two coordinate systems to cover the interior of a Maclaurin spheroid. The coordinates $(\zeta, \mu, \phi)$ are defined by 
\begin{equation} 
x = a (1+\zeta^2)^{1/2} (1-\mu^2)^{1/2} \cos\phi, \qquad 
y = a (1+\zeta^2)^{1/2}(1-\mu^2)^{1/2} \sin\phi, \qquad 
z = a\, \zeta \mu, 
\label{coords_2} 
\end{equation} 
with $0 \leq \zeta \leq \zeta_0$, $-1 \leq \mu \leq 1$, and $0 \leq \phi < 2\pi$. Surfaces of constant $\zeta$ are spheroids, while surfaces of constant $\mu$ are hyperboloids. The specific enthalpy of Eq.~(\ref{enthalpy}) becomes 
\begin{equation} 
h = h_c \biggl[ 1 - \frac{(1+\zeta^2)(1-\mu^2)}{1 + \zeta_0^2} - \frac{\zeta^2 \mu^2}{\zeta_0^2} \biggr] 
\label{h_coords2} 
\end{equation} 
in the new coordinates. The surface of the Maclaurin spheroid is described by $\zeta = \zeta_0$. 

The coordinates $(\xi,\nu,\phi)$ are defined by 
\begin{equation} 
x = b (1-\xi^2)^{1/2}(1-\nu^2)^{1/2} \cos\phi, \qquad 
y = b (1-\xi^2)^{1/2}(1-\nu^2)^{1/2} \sin\phi, \qquad 
z = b \frac{\sqrt{4-\kappa^2}}{\kappa}\, \xi\nu, 
\label{coords_3} 
\end{equation} 
where $\kappa$ is a parameter in the interval $-2 < \kappa < 2$, and 
\begin{equation} 
b := \frac{\sqrt{4(1+\zeta_0^2) - \kappa^2}}{\sqrt{4 - \kappa^2}}\, a. 
\label{b_def} 
\end{equation} 
When $\kappa > 0$, Lindblom and Ipser \cite{lindblom-ipser:99} show that to cover the entire interior of the Maclaurin spheroid, the coordinates must lie in the intervals $\xi_0 \leq \xi \leq 1$, $-\xi_0 \leq \nu \leq \xi_0$, and $0 \leq \phi < 2\pi$, with 
\begin{equation} 
\xi_0 := \frac{\kappa \zeta_0}{\sqrt{4(1+\zeta_0^2) - \kappa^2}}. 
\label{xi0_def} 
\end{equation} 
When $\kappa < 0$, so that $\xi_0 < 0$, the appropriate ranges are $-1 \leq \xi \leq \xi_0$, $\xi_0 \leq \nu \leq -\xi_0$, and $0 \leq \phi < 2\pi$. It is useful to note that $\xi_0$ is bounded by $|\xi_0| < \frac{1}{2} |\kappa| < 1$; the upper bound is reached in the limit $\zeta_0 \to \infty$. Surfaces of constant $\xi$ are spheroids, and so are the surfaces of constant $\nu$.  

The specific enthalpy of Eq.~(\ref{enthalpy}) takes the form 
\begin{equation} 
h = h_c \biggl[ 1 - \frac{(1-\xi^2)(1-\nu^2)}{1 - \xi_0^2} - \frac{\xi^2\nu^2}{\xi_0^2} \biggr] 
\label{h_coords3} 
\end{equation} 
in these coordinates. The description of the surface is complicated: it is broken up into three segments, in which either $\xi$ or $\nu$ is equal to $\xi_0$; the details are presented in Ref.~\cite{lindblom-ipser:99}. 

The volume element is 
\begin{equation} 
d{\cal V} = b^3 \frac{\sqrt{4-\kappa^2}}{|\kappa|}\, (\xi^2 - \nu^2)\, d\xi d\nu d\phi
\label{dV_coord3} 
\end{equation} 
in the $(\xi,\nu,\phi)$ coordinates. Taking into account the distinct coordinate intervals for $\kappa > 0$ and $\kappa < 0$, we find that the volume of the Maclaurin spheroid is given by 
\begin{equation} 
{\cal V} = \frac{4\pi}{3} \frac{\sqrt{4-\kappa^2}}{\kappa} \xi_0 (1 - \xi_0^2)\, b^3
= \frac{4\pi}{3} \zeta_0 (1 + \zeta_0^2)\, a^3 = \frac{4\pi}{3} R^3,
\end{equation} 
as it should be.   

\subsection{Fluid perturbations} 
\label{sec:fluid}  
 
We consider a star of mass $M$ and mean radius $R$, rotating rigidly with an angular velocity $\Omega$. We take the star to be described by a perfect fluid with a barotropic equation of state $\rho = \rho(h)$, $p = p(h)$, where $\rho$ is the mass density, $p$ the pressure, and $h$ the specific enthalpy, defined by $dh = \rho^{-1}\, dp$. The physics of the fluid is governed by Euler's equation, the continuity equation, and Poisson's equation for the gravitational potential. 

Euler's equation is 
\begin{equation} 
\partial_t v_a + v^b \nabla_b v_a + \nabla_a V = 0, 
\label{euler} 
\end{equation} 
where $v^a$ is the velocity field and 
\begin{equation} 
V := h - U 
\label{Vdef} 
\end{equation} 
is the hydrodynamic potential, with $U$ denoting the gravitational potential. The equation is written in covariant form, and can therefore be formulated in any coordinate system; the same remark applies to most equations in this section\footnote{Equations that are formulated explicitly in Cartesian coordinates use tensor indices $j$, $k$, $p$, and so on, from the second half of the Latin alphabet. Covariant equations feature indices $a$, $b$, $c$, and so on, from the first half of the alphabet.}. The continuity equation is 
\begin{equation} 
\partial_t \rho + \nabla_a (\rho v^a) = 0, 
\label{continuity} 
\end{equation} 
and Poisson's equation is 
\begin{equation} 
\nabla^2 U +  4\pi G \rho = 0. 
\label{poisson} 
\end{equation}. 

The unperturbed fluid is in a state of rigid rotation, with an angular-velocity vector $\Omega^a$ of constant magnitude $\Omega$. In Cartesian coordinates we have that $\Omega^j = (0,0, \Omega)$; the vector is aligned with the $z$-axis. The unperturbed velocity vector is $v^a = \Omega \phi^a$, where $\phi^a$ is the azimuthal Killing vector. In Cartesian coordinates we have that $v_j = \varepsilon_{jkp} \Omega^k x^p = (-\Omega y, \Omega x, 0)$. It follows that $\partial_k v_j = \varepsilon_{jpk} \Omega^p$, and the covariant version of this is 
\begin{equation} 
\nabla_a v_b = \varepsilon_{abc} \Omega^c.  
\label{gradv} 
\end{equation} 

Next we consider a fluid perturbation described by the Eulerian changes $\{ \delta v^a, \delta\rho, \delta p, \delta h, \delta U \}$. The perturbation of Euler's equation is 
\begin{equation} 
\partial_t \delta v_a + \delta v^b\, \nabla_b v_a + v^b \nabla_b \delta v_a 
+ \nabla_a \delta V = 0.  
\end{equation} 
With Eq.~(\ref{gradv}) and some manipulations, this can be put in the useful form 
\begin{equation} 
(\partial_t + \Lie_v) \delta v_a + 2\varepsilon_{abc} \Omega^b \delta v^c 
+ \nabla_a \delta V = 0, 
\label{dEuler} 
\end{equation} 
where $\Lie_v \delta v_a = v^b \nabla_b \delta v_a + \delta v_b \nabla_a v^b$ is the Lie derivative of the velocity perturbation in the direction of the unperturbed velocity.  

The velocity perturbation can be related to a Lagrangian displacement vector $\Xi^a$ by \cite{friedman-schutz:78a} 
\begin{equation} 
\delta v^a = \partial_t \Xi^a - \Lie_\Xi v^a = (\partial_t + \Lie_v) \Xi^a. 
\label{xi_vs_dv} 
\end{equation}  
The perturbed continuity equation then implies that 
\begin{equation} 
\delta \rho = -\nabla_a (\rho \Xi^a), 
\label{drho} 
\end{equation} 
and $\delta h$ can be obtained from the equation of state. The perturbed Poisson equation is 
\begin{equation} 
\nabla^2 \delta U + 4\pi G \delta\rho = 0, 
\label{dPoisson} 
\end{equation} 
and it determines $\delta U$. 

The perturbation equations come with the requirement that $\Delta h = 0$ at the stellar surface, where $\Delta h = \delta h + \Xi^a \nabla_a h$ is the Lagrangian change in the specific enthalpy. We therefore have 
\begin{equation} 
\delta h + \Xi^a \nabla_a h = 0 \qquad (h = 0).  
\label{Dh}
\end{equation}

The operator $\partial_t + \Lie_v$ plays an important role in the subsequent developments. To elucidate its meaning, we first point out that in any coordinates $(x^1, x^2, \phi)$ in which $v^\phi = \Omega$ is the only nonvanishing component of the velocity vector (with $\phi$ denoting the azimuthal angle around the rotation axis), $\Lie_v f = \Omega \partial_\phi f$ for any tensor $f$ (with any number of indices in any position). So we have that $(\partial_t + \Lie_v) f = (\partial_t + \Omega \partial_\phi) f$ in such a coordinate system. This can readily be interpreted as a partial time derivative in the star's corotating frame. To see this, let $\phi$ continue to be the angle as measured in the nonrotating frame, and let $\varphi$ be the angle in the corotating frame; the coordinates are related by $\varphi = \phi - \Omega t$. Let $f$ be a function of $t$ and $\phi$, as well as the two other coordinates. The operator $\partial_t$ is a partial derivative with respect to $t$ at constant $\phi$. Let $\partial'_t$ be the partial derivative at constant $\varphi$. Then writing $f(t,\phi) = f(t,\varphi + \Omega t)$, it is immediately clear that $\partial'_t f = \partial_t f + \Omega \partial_\phi f = (\partial_t + \Lie_v)f$. This implies that as claimed, $\partial_t + \Lie_v$ is a partial time derivative in the corotating frame.   

\subsection{Two-potential formalism} 
\label{subsec:2pot} 

We wish to solve the perturbation equations to obtain the star's inertial modes of vibration. To achieve this we exploit the two-potential formalism of Ipser and Lindblom \cite{ipser-lindblom:90}; the potentials are $\delta V$ and $\delta U$.  

We take all perturbation variables $f$ to be proportional to $e^{im\phi} e^{-i\omega t}$, where $m$ is an integer and $\omega$ is the mode frequency in the nonrotating frame. It follows that 
$(\partial_t + \Lie_v)f = -i\sigma f$, where
\begin{equation}
\sigma := \omega - m \Omega
\label{sigma_def}
\end{equation}
is the mode frequency in the corotating frame; the interpretation follows at once from the identity $e^{im\phi} e^{-i\omega t} = e^{im\varphi} e^{-i\sigma t}$. We therefore have that 
\begin{equation}
\Xi^a = \frac{i}{\sigma}\, \delta v^a,
\label{xi_dv} 
\end{equation}
and the continuity equation of Eq.~(\ref{drho}) can be re-expressed as
\begin{equation}
\delta \rho = -\frac{i}{\sigma} \nabla_a (\rho \delta v^a).
\label{drho_dv}
\end{equation}
The Euler equation (\ref{dEuler}) becomes
\begin{equation}
(\sigma g_{ab} - 2i \varepsilon_{abc} \Omega^c) \delta v^b = -i \nabla_a \delta V. 
\label{dEuler2} 
\end{equation}
The tensor multiplying $\delta v^b$ can be inverted, and we get 
\begin{equation}
\delta v^a = -i Q^{ab} \nabla_b \delta V 
\label{dv_dV}
\end{equation}
with
\begin{equation}
Q^{ab} := \frac{1}{\sigma^2 - 4\Omega^2} \biggl( \sigma g^{ab}
- \frac{4}{\sigma} \Omega^a \Omega^b + 2i \varepsilon^{abc} \Omega_c \biggr).
\label{Q_def}
\end{equation} 
Equation (\ref{dv_dV}) is the key equation of the two-potential formalism. 

To obtain the final form of the mode equations we insert Eq.~(\ref{dv_dV}) within Eq.~(\ref{drho_dv}) and find
\begin{equation}
\delta \rho = -\frac{1}{\sigma}\nabla_a \bigl( \rho Q^{ab} \nabla_b \delta V \bigr).
\end{equation}
On the other hand, the barotropic equation of state $\rho = \rho(h)$ implies that $\delta \rho = (d\rho/dh)\, \delta h = (d\rho/dh) (\delta V + \delta U)$. Equating the results, we have that
\begin{equation}
\nabla_a \bigl( \rho Q^{ab} \nabla_b \delta V \bigr)
+ \sigma \frac{d\rho}{dh} (\delta V + \delta U) = 0.
\label{mode_eq1}
\end{equation}
This is the first mode equation. The second mode equation is a restatement of Eq.~(\ref{dPoisson}),
\begin{equation}
\nabla^2 \delta U + 4\pi G \frac{d\rho}{dh} (\delta V + \delta U) = 0.
\label{mode_eq2}
\end{equation}
These equations come with the surface condition of Eq.~(\ref{Dh}), or
\begin{equation}
Q^{ab} (\nabla_a h) (\nabla_b \delta V) + \sigma(\delta V + \delta U) = 0 \qquad (h=0)
\label{mode_eq3}
\end{equation}
after inserting Eqs.~(\ref{xi_dv}) and (\ref{dv_dV}).

Once the mode equations are integrated for $\delta V$ and $\delta U$, the velocity field can be recovered from Eq.~(\ref{dv_dV}), and the Lagrangian displacement from Eq.~(\ref{xi_dv}). The perturbation in the specific enthalpy is $\delta h = \delta V + \delta U$, and $\delta \rho$, $\delta p$ can be recovered from the equation of state. Equations (\ref{mode_eq1}), (\ref{mode_eq2}), and (\ref{mode_eq3}) constitute an eigenvalue problem for the mode frequencies $\sigma$ and the mode functions $\Xi^a$.

\subsection{Specialization to a Maclaurin spheroid}
\label{subsec:maclaurin}

The equation of state appropriate to a Maclaurin spheroid is Eq.~(\ref{EOS}), from which it follows that $d\rho/dh = \rho_0 \delta(h)$ and $\nabla_a \rho = \rho_0 \delta(h) \nabla_a h$. Making the substitutions in Eq.~(\ref{mode_eq1}) returns
\begin{equation} 
Q^{ab} \nabla_{ab} \delta V = 0
\label{mac1_tmp} 
\end{equation} 
and 
\begin{equation} 
Q^{ab} (\nabla_a h) (\nabla_b \delta V) + \sigma(\delta V + \delta U) = 0 \qquad (h=0). 
\label{mac2} 
\end{equation} 
Equation (\ref{mac2}) is identical to the surface condition of Eq.~(\ref{mode_eq3}). Using Eq.~(\ref{Q_def}), Eq.~(\ref{mac1_tmp}) becomes
\begin{equation} 
\bigl( \sigma^2 g^{ab} - 4 \Omega^a \Omega^b \bigr) \nabla_{ab} \delta V = 0. 
\label{mac1} 
\end{equation} 

Equation (\ref{mode_eq2}) reduces to 
\begin{equation} 
\nabla^2 \delta U + 4\pi G \rho_0 (\delta V + \delta U) \delta(h) = 0. 
\label{mac3_tmp} 
\end{equation} 
This gives rise to two equations, one for the interior and exterior of the Maclaurin spheroid, 
\begin{equation} 
\nabla^2 \delta U = 0, 
\label{mac3} 
\end{equation} 
and one surface condition. This is obtained by writing $\delta U = \delta U_{\rm in} \Theta(h) + \delta U_{\rm out} \Theta(-h)$, where $\delta U_{\rm in}$ and $\delta U_{\rm out}$ are respectively the interior and exterior solutions to Eq.~(\ref{mac3}). We require continuity at $h=0$, so that $\delta U_{\rm in}(h=0) = \delta U_{\rm out}(h=0)$. Substitution within Eq.~(\ref{mac3_tmp}) then produces 
\begin{equation} 
\nabla^a h \bigl( \nabla_a \delta U_{\rm out} - \nabla_a \delta U_{\rm in} \bigr) 
- 4\pi G \rho (\delta V + \delta U) = 0 \qquad (h = 0). 
\label{mac4} 
\end{equation} 
The normal derivative of the gravitational potential is discontinuous by virtue of the discontinuity in the density. Notice that we no longer distinguish between $\rho$ and $\rho_0$ in Eq.~(\ref{mac4}).     

\subsection{Inertial modes} 
\label{sec:inertial} 

We integrate Eqs.~(\ref{mac1}) and (\ref{mac3}) for the potentials $\delta V$ and $\delta U$. This can be done for any value of the mode frequency $\sigma$. The eigenfrequencies are obtained in a second stage, by imposing the surface conditions of Eqs.~(\ref{mac2}) and (\ref{mac4}). It is understood that the potentials are proportional to $e^{-i\omega t}$; this factor, however, is omitted from all equations below. 

We adopt the following notation for the associated Legendre functions. When the argument $x$ is restricted to the interval $-1 \leq x \leq 1$, we denote the functions ${\sf P}_\ell^m(x)$ and ${\sf Q}_\ell^m(x)$; when the argument $z$ is in the interval $1 \leq z < \infty$, or when it is complex, they are denoted $P_\ell^m(z)$ and $Q_\ell^m(z)$. All definitions are supplied in Sec.~14.3 of the {\it NIST Handbook of Mathematical Functions} \cite{NIST:10}, hereafter referred to as ``NIST''.  With an appropriate choice of branch cut, these definitions agree with the implementation provided by the Maple symbolic manipulation software, which was used to perform calculations. Initially we take $\ell$ and $m$ to be integers restricted by $\ell \geq 0$ and $-\ell \leq m \leq \ell$; their values will be further restricted below.      

The gravitational potential $\delta U$ must be a solution to Laplace's equation (\ref{mac3}), which we formulate in the $(\zeta, \mu, \phi)$ coordinates of Sec.~\ref{sec:coordinates}. The appropriate solutions are 
\begin{equation} 
\delta U_{\rm in} = \alpha  \frac{P_\ell^m(i\zeta)}{P_\ell^m(i\zeta_0)} {\sf P}_\ell^m(\mu) e^{im\phi} 
\label{U_in} 
\end{equation} 
for the interior potential, and   
\begin{equation} 
\delta U_{\rm out} = \alpha \frac{iQ_\ell^m(i\zeta)}{iQ_\ell^m(i\zeta_0)} {\sf P}_\ell^m(\mu) e^{im\phi} 
\label{U_out} 
\end{equation} 
for the exterior potential, where $\alpha$ is a normalization constant. The modes are labeled with the integers $\ell$ and $m$, and with a third integer $n$ that sequences the distinct eigenfrequencies for each $\ell$ and $m$. The functions $P_\ell^m(i\zeta)$ and $iQ_\ell^m(i\zeta)$ are real for a real $\zeta$. 

Next we turn to Eq.~(\ref{mac1}) for the hydrodynamic potential $\delta V$. This equation is formulated in the $(\xi,\nu,\phi)$ coordinates of Sec.~\ref{sec:coordinates}, in which we set 
\begin{equation} 
\kappa = \sigma/\Omega. 
\label{kappa_vs_sigma} 
\end{equation} 
The appropriate solutions are 
\begin{equation} 
\delta V = \beta  \frac{{\sf P}_\ell^m(\xi)}{{\sf P}_\ell^m(\xi_0)} {\sf P}_\ell^m(\nu) e^{im\phi}, 
\label{V_sol} 
\end{equation} 
where $\beta$ is a constant. 

Equations (\ref{U_in}), (\ref{U_out}), and (\ref{V_sol}) provide solutions to Eqs.~(\ref{mac1}) and (\ref{mac3}) for the two potentials. To get the eigenvalues, we must also impose the surface conditions of Eq.~(\ref{mac2}) and (\ref{mac4}). The detailed analysis is presented in Lindblom and Ipser \cite{lindblom-ipser:99}. The outcome is that $\kappa$ is determined by the eigenvalue equation 
\begin{equation} 
AB - BC + AD = 0,  
\label{eigen} 
\end{equation} 
where 
\begin{subequations} 
\label{ABCD} 
\begin{align} 
A &:= \frac{1 + \zeta_0^2}{\sqrt{4(1+\zeta_0^2) - \kappa^2}} \frac{1}{{\sf P}_\ell^m(\xi_0)} 
\frac{d {\sf P}_\ell^m(\xi_0)}{d\xi_0} + \frac{2m \zeta_0}{4-\kappa^2}, \\
B &:= (-1)^{m+1} \frac{(\ell+m)!}{(\ell-m)!} \frac{1}{ P_\ell^m(i\zeta_0)\, iQ_\ell^m(i\zeta_0) }, \\ 
C &:= \frac{(1 + 3\zeta_0^2)\arccot\zeta_0 - 3\zeta_0}{2(1-\zeta_0\, \arccot\zeta_0)} \kappa, \\ 
D &:= \frac{1}{\zeta_0(1 - \zeta_0\, \arccot\zeta_0)}. 
\end{align} 
\end{subequations} 
Once a solution is obtained, the ratio $\alpha/\beta$ is given by $(A-C)/C$, and the mode functions are fully determined up to a normalization constant. We recall that $\xi_0$ is related to $\kappa$ by Eq.~(\ref{xi0_def}). We recall also that the frequency in the corotating frame is $\sigma = \kappa \Omega$; in the nonrotating frame we have $\omega = \sigma + m\Omega = (\kappa + m)\Omega$. 

The eigenfrequencies are computed by subjecting Eq.~(\ref{eigen}) to a root-finding algorithm. We use the built-in numerical routine provided by Maple, and make sure to employ a sufficient number of significant digits (30 does the trick) to obtain a reliable evaluation of the Legendre functions. 

The associated Legendre functions satisfy the identities [NIST's Eqs.~(14.7.17), (14.9.13), and (14.9.14)]
\begin{equation} 
{\sf P}_\ell^m(-x) = (-1)^{\ell+m} {\sf P}_\ell^m(x), \qquad 
P_\ell^{-m}(z) = \frac{(\ell-m)!}{(\ell+m)!} P_\ell^m(z), \qquad 
Q_\ell^{-m}(z) = \frac{(\ell-m)!}{(\ell+m)!} Q_\ell^m(z). 
\end{equation} 
It follows from this that under the mapping $m \to -m$, $\kappa \to -\kappa$ and $\xi_0 \to -\xi_0$, the quantities listed in Eq.~(\ref{ABCD}) change according to $A \to -A$, $B \to B$, $C \to -C$, and $D \to D$. The eigenvalue equation (\ref{eigen}) is therefore invariant under the mapping. Sending $m$ to $-m$ therefore sends an eigenvalue $\kappa$ to $-\kappa$; the frequency in the rotating frame changes sign, and the frequency in the nonrotating frame does as well. A mode with a negative value of $m$ is therefore the complex conjugate of a mode with the corresponding positive value. For this reason it is sufficient to restrict our attention to $m \geq 0$. 

Lindblom and Ipser \cite{lindblom-ipser:99} show that there are no nontrivial solutions to Eq.~(\ref{eigen}) when $\ell=0$ and $\ell=1$. For $\ell \geq 2$, we have that $m$ must be smaller than or equal to $\ell-1$. The number of distinct solutions to Eq.~(\ref{eigen}) is then equal to $\ell - m$ when $m \neq 0$, and $\ell - 1$ when $m = 0$. The modes with $m = \ell -1$ constitute the subclass of $r$-modes. 

\subsection{$r$-mode with $\ell = 2$ and $m=1$} 
\label{subsec:zerof} 

It is interesting to note that the $r$-mode with $\ell = 2$ and $m=1$ possesses the eigenvalue $\kappa = -1$, or 
\begin{equation} 
\sigma = -\Omega,  
\end{equation} 
for any value of $\zeta_0$ (and therefore $\Omega$). The mode frequency in the nonrotating frame is $\omega = 0$. If we choose the normalization $\delta V = {\sf P}_2^1(\xi) {\sf P}_2^1(\nu) e^{i\phi}$ for the hydrodynamic potential, then 
\begin{equation} 
\delta V = 9 \xi (1-\xi^2)^{1/2} \nu (1-\nu^2)^{1/2} e^{i\phi} = -\frac{3\sqrt{3}}{b^2} (x+iy) z  
\end{equation}
with $b = (1 + \frac{4}{3} \zeta_0^2)^{1/2} a$. It follows from this and Eqs.~(\ref{xi_dv}), (\ref{dv_dV}), and (\ref{Q_def}) that up to a normalization factor, the mode's Lagrangian displacement vector is given by 
\begin{equation} 
\Xi^j \propto (z, iz, -x - i y) 
\end{equation} 
in Cartesian coordinates. 

\section{Tidal driving of inertial modes} 
\label{sec:driving} 

In this section we describe how the inertial modes constructed in Sec.~\ref{sec:inertial} are driven by Newtonian (gravitoelectric) and post-Newtonian (gravitomagnetic) tidal fields. We rely heavily on the methods devised by Schenk {\it et al.}\ \cite{schenk-etal:01} to represent a driven perturbation of a rotating star as a sum over normal modes.  

\subsection{Mode equations} 

It is cleaner to formulate the fluid equations of Sec.~\ref{sec:fluid} in the star's corotating frame, which rotates with an angular velocity $\Omega$ with respect to the original, nonrotating frame. If we also add an external force density $f_a$ on the right-hand side of Eq.~(\ref{dEuler}), it becomes 
\begin{equation} 
\partial_t \delta v_a + 2 \varepsilon_{abc} \Omega^b \delta v^c + \nabla_a \delta V = f_a, 
\end{equation} 
where $\partial_t$ (previously $\partial_t + \Lie_v$) denotes the time derivative operator in the corotating frame --- recall the discussion in the last paragraph of Sec.~\ref{sec:fluid}. The mode formalism makes extensive use of the Lagrangian displacement vector, which is defined by Eq.~(\ref{xi_vs_dv}); in the new notation this is $\delta v^a = \partial_t \Xi^a$. Making the substitution, Euler's equation becomes 
\begin{equation} 
\partial_{tt} \delta \Xi_a + 2 \varepsilon_{abc} \Omega^b \partial_t \Xi^c + \nabla_a \delta V = f_a. 
\label{dEuler_xi} 
\end{equation} 
The remaining fluid equations are Eq.~(\ref{drho}) for $\delta \rho$, Eq.~(\ref{dPoisson}) for the gravitational potential $\delta U$, the surface condition of Eq.~(\ref{Dh}), and the definition $\delta V = \delta h - \delta U$ for the hydrodynamic potential. 

It is useful to introduce a linear-algebra notation in which $\Xi^a$ is mapped to an abstract vector $\Xi$ in a Hilbert space, the operation $\KK^a_{\ b} \Xi^b := \varepsilon^a_{\ bc} \Omega^b \Xi^c$ is mapped to $\KK \Xi$, and the external force $f^a$ is mapped to $f$. We also introduce an inner product $\bkt{\eta}{\Xi}$ between two vectors $\eta$ and $\Xi$, defined by 
\begin{equation}
\bkt{\eta}{\Xi} := \int_V \rho\, \bar{\eta}_a \Xi^a\, d{\cal V}, 
\end{equation}
where an overbar indicates complex conjugation. 

We wish to represent $\Xi$, the solution to Eq.~(\ref{dEuler_xi}), as a sum over normal modes. We let each mode be described by the Lagrangian displacement 
\begin{equation} 
\Xi_K(t,x^a) = \chi_K(x^a) e^{-i\sigma_K t}, 
\end{equation} 
where $K$ is a mode label and $\sigma_K$ the mode frequency in the corotating frame --- the frequency in the nonrotating frame is $\omega_K = \sigma_K + m\Omega$. We follow Schenk {\it et al.}\ \cite{schenk-etal:01} and adopt a ``phase-space'' representation for $\Xi$, according to which  
\begin{equation} 
\Xi  = \sum_K q_K(t)\, \chi_K, \qquad  
\partial_t \Xi = \sum_K (-i\sigma_K) q_K(t)\, \chi_K, 
\label{mode-sum} 
\end{equation} 
for some amplitudes $q_K(t)$. Equation (\ref{dEuler_xi}) implies that these are determined by 
\begin{equation} 
\dot{q}_K + i\sigma_K q_K = -\frac{\bkt{\chi_K}{f}}{2i\sigma_K N_K}, 
\label{q_eq} 
\end{equation} 
where 
\begin{equation} 
N_K := \bkt{\chi_K}{\chi_K} - \frac{1}{i\sigma_K} \bkt{\chi_K}{\KK \chi_K} 
\label{Ndef} 
\end{equation} 
provides a notion of mode norm. (This is to be interpreted with caution: while $N_K$ is real, it is not positive-definite.) 

The modes that interest us here are the inertial modes of Sec.~\ref{sec:inertial}, and the mode label $K$ therefore includes the integers $\ell$ and $m$. It also includes a third integer $n$, which sequences the distinct modes with a given value of $\ell$ and $m$. 

The amplitudes $q_K$ depend on a choice of normalization for the modes. To remove this arbitrariness it is convenient to normalize the modes so that $|N_K| = 1$. Equivalently, we can introduce normalized modes and overlap integrals defined by 
\begin{equation} 
\chi^{\rm norm}_K := \frac{\chi_K}{\sqrt{|N_K|}}, \qquad 
\bkt{\chi_K}{f}^{\rm norm} := \frac{\bkt{\chi_K}{f}}{\sqrt{|N_K|}}. 
\end{equation} 
The corresponding normalized amplitudes are then $q^{\rm norm}_K = \sqrt{|N_K|} q_K$. 

\subsection{Tidal moments}     

We place the Maclaurin spheroid within a generic tidal environment characterized by a Newtonian (gravitoelectric) quadrupole tidal moment ${\cal E}_{jk}(t)$, and a post-Newtonian (gravitomagnetic) quadrupole tidal moment ${\cal B}_{jk}(t)$ --- these symmetric-tracefree tensors are given in Cartesian coordinates, in the corotating frame. We shall calculate the tidal potentials associated with these moments, obtain the corresponding force densities, and compute the overlap integrals with the spheroid's inertial modes.  

To perform these calculations it is convenient to introduce the Fourier decompositions
\begin{equation} 
{\cal E}_{jk}(t) = \int_{-\infty}^\infty \tilde{\cal E}_{jk}(\sigma_{\rm ext}) e^{-i\sigma_{\rm ext} t}\, dt, \qquad 
{\cal B}_{jk}(t) = \int_{-\infty}^\infty \tilde{\cal B}_{jk}(\sigma_{\rm ext}) e^{-i\sigma_{\rm ext} t}\, dt, 
\end{equation} 
where $\sigma_{\rm ext}$ is the external frequency (as opposed to mode frequency) in the corotating frame. Because the tidal moments are real, their Fourier transforms satisfy
\begin{equation} 
\tilde{\cal E}_{jk}(-\sigma_{\rm ext}) = \tilde{\cal E}^*_{jk}(\sigma_{\rm ext}), \qquad 
\tilde{\cal B}_{jk}(-\sigma_{\rm ext}) = \tilde{\cal B}^*_{jk}(\sigma_{\rm ext}),
\end{equation} 
where an asterisk (like the overbar previously) indicates complex conjugation. 

We also perform a decomposition of the tidal moments in a tensor basis that projects out each one of their $m$-components; this step is crucial in the calculation of overlap integrals, which are specific to each value of the mode label $m$. This is accomplished with the help of the symmetric-tracefree tensors (see Box~1.5 of Ref.~\cite{poisson-will:14}) 
\begin{equation} 
({\scrpt Y}_2^{\pm 2})^{jk} = \frac{1}{8} \sqrt{\frac{30}{\pi}} \left( 
\begin{array}{ccc} 
1 & \mp i & 0 \\ 
\mp i & -1 & 0 \\ 
0 & 0 & 0 
\end{array} \right), \qquad 
({\scrpt Y}_2^{\pm 1})^{jk} = \mp \frac{1}{8} \sqrt{\frac{30}{\pi}} \left( 
\begin{array}{ccc} 
0 & 0 & 1 \\ 
0 & 0 & \mp i \\ 
1 & \mp i & 0 
\end{array} \right), \qquad 
({\scrpt Y}_2^{0})^{jk} = -\frac{1}{4} \sqrt{\frac{5}{\pi}} \left( 
\begin{array}{ccc} 
1 & 0 & 0 \\ 
0 & 1 & 0 \\ 
0 & 0 & -2 
\end{array} \right), 
\label{STFtensors} 
\end{equation}   
which satisfy $({\scrpt Y}_2^{-m})^{jk} = (-1)^m (\bar{\scrpt Y}_2^{m})^{jk}$. These tensors are defined so that the spherical harmonics of degree $\ell = 2$ can be expressed as [Eq.~(1.167) of Ref.~\cite{poisson-will:14}] 
\begin{equation} 
Y_2^m(\theta,\phi) =  (\bar{\scrpt Y}_2^{m})_{jk}\, n^j n^k, 
\end{equation} 
where $n^j = (\sin\theta\cos\phi, \sin\theta\sin\phi, \cos\theta)$ is the radial unit vector. 

The decomposition of the tidal moments is accomplished by 
\begin{equation} 
\tilde{\cal E}_{jk} = \sum_{m=-2}^2 \tilde{\cal E}^m (\bar{\scrpt Y}_2^{m})_{jk}, \qquad 
\tilde{\cal B}_{jk} = \sum_{m=-2}^2 \tilde{\cal B}^m (\bar{\scrpt Y}_2^{m})_{jk} 
\label{EB_decomposition} 
\end{equation} 
with 
\begin{equation} 
\tilde{\cal E}^m := \frac{8\pi}{15} ({\scrpt Y}_2^{m})^{jk}\, \tilde{\cal E}_{jk}, \qquad 
\tilde{\cal B}^m := \frac{8\pi}{15} ({\scrpt Y}_2^{m})^{jk}\, \tilde{\cal B}_{jk}. 
\label{EB_projections} 
\end{equation} 
The projections satisfy the reality conditions 
\begin{equation} 
\tilde{\cal E}^{-m}(\sigma_{\rm ext}) = (-1)^m \bigl[ \tilde{\cal E}^m(-\sigma_{\rm ext}) \bigr]^*, \qquad 
\tilde{\cal B}^{-m}(\sigma_{\rm ext}) = (-1)^m \bigl[ \tilde{\cal B}^m(-\sigma_{\rm ext}) \bigr]^*. 
\end{equation} 
Because of this redundancy, it will be sufficient below to calculate overlap integrals for $m = \{0, 1 ,2\}$ only. 

As specific examples of tidal moments, we take those produced by a companion body of mass $M'$ moving on a circular orbit of radius $r$ and angular velocity $\varpi$, with 
\begin{equation} 
\varpi = \sqrt{\frac{G M_{\rm tot}}{r^3}}, 
\label{orbital_frequency} 
\end{equation} 
where $M_{\rm tot} := M + M'$ is the binary's total mass. The companion's position with respect to the spheroid's center of mass is in the direction of the unit vector $\bm{n}$, and the normal to the orbital plane points along the unit vector $\bm{l}$. We give the orbit a generic orientation, so that it possesses an inclination angle $\iota$ with respect to the spheroid's equatorial plane. In the nonrotating frame the orbital vectors are given by [Eqs.~(3.42) and (3.45) of Ref.~\cite{poisson-will:14}] 
\begin{subequations} 
\label{orbit_vectors} 
\begin{align} 
\bm{n} &= \bigl( \cos\alpha\, \cos\varpi t - \cos\iota\, \sin\alpha\, \sin\varpi t, 
\sin\alpha\, \cos\varpi t + \cos\iota\, \cos\alpha\, \sin\varpi t, 
\sin\iota\, \sin\varpi t \bigr), \\
\bm{l} &= \bigl( \sin\iota\, \sin\alpha, -\sin\iota\, \cos\alpha, \cos\iota \bigr), 
\end{align} 
\end{subequations} 
where $\alpha$ is the ``longitude of the ascending node'' (traditionally denoted $\Omega$, which is reserved here for the spheroid's angular velocity); a helpful diagram is provided by Fig.~3.2 of Ref.~\cite{poisson-will:14}. While the line of nodes is fixed in the nonrotating frame, it rotates with an angular velocity $-\Omega$ in the corotating frame, and the transformation of $\bm{n}$ and $\bm{l}$ to the corotating frame is effected by letting $\alpha \to \alpha - \Omega t$. After the transformation is accomplished, $\alpha$ has served its purpose and can be set equal to zero without loss of generality. The upshot is that Eqs.~(\ref{orbit_vectors}) with $\alpha = -\Omega t$ give the orbital vectors in the corotating frame. 

The tidal moments produced by the companion body are given by \cite{taylor-poisson:08} 
\begin{equation} 
{\cal E}_{jk} = -\frac{GM'}{r^3} \bigl( 3 n_j n_k - \delta_{jk} \bigr), \qquad 
{\cal B}_{jk} = \frac{3GM'v'}{r^3} \bigl( l_j n_k + n_k l_k \bigr), 
\end{equation} 
where $v' = r\varpi$ is the orbital velocity. Inserting these in Eq.~(\ref{EB_projections}) returns 
\begin{subequations} 
\label{E_binary} 
\begin{align} 
\tilde{\cal E}^{m=2} &= -\frac{\sqrt{30\pi}}{20} \frac{GM'}{r^3} \Bigl[ 
2\sin^2\iota\, \delta(\sigma_{\rm ext}+2\Omega) 
+ (1+\cos\iota)^2\, \delta(\sigma_{\rm ext}+2\Omega-2\varpi) 
+ (1-\cos\iota)^2\, \delta(\sigma_{\rm ext}+2\Omega+2\varpi) \Bigr], \\ 
\tilde{\cal E}^{m=1} &= i\frac{\sqrt{30\pi}}{10} \frac{GM'}{r^3} \Bigl[
-2\sin\iota\cos\iota\, \delta(\sigma_{\rm ext}+\Omega)  
+ \sin\iota(1+\cos\iota)\, \delta(\sigma_{\rm ext}+\Omega-2\varpi) 
- \sin\iota(1-\cos\iota)\, \delta(\sigma_{\rm ext}+\Omega+2\varpi) \Bigr], \\ 
\tilde{\cal E}^{m=0} &= \frac{\sqrt{5\pi}}{10} \frac{GM'}{r^3} \Bigl[
3\sin^2\iota\, \delta(\sigma_{\rm ext}-2\varpi) 
+ 3\sin^2\iota\, \delta(\sigma_{\rm ext}+2\varpi) 
- 2(1 - 3\cos^2\iota)\, \delta(\sigma_{\rm ext}) \Bigr]
\end{align} 
\end{subequations} 
and 
\begin{subequations} 
\label{B_binary} 
\begin{align} 
\tilde{\cal B}^{m=2} &= i\frac{\sqrt{30\pi}}{5} \frac{GM' v'}{r^3} \Bigl[ 
\sin\iota(1 + \cos\iota)\, \delta(\sigma_{\rm ext} + 2\Omega - \varpi) 
+ \sin\iota(1 - \cos\iota)\, \delta(\sigma_{\rm ext} + 2\Omega + \varpi) \Bigr], \\ 
\tilde{\cal B}^{m=1} &= \frac{\sqrt{30\pi}}{5} \frac{GM' v'}{r^3} \Bigl[ 
 (1+\cos\iota)(1-2\cos\iota)\, \delta(\sigma_{\rm ext}+\Omega-\varpi) 
- (1-\cos\iota)(1+2\cos\iota)\, \delta(\sigma_{\rm ext}+\Omega+\varpi) \Bigr], \\ 
\tilde{\cal B}^{m=0} &= i\frac{6\sqrt{5\pi}}{5} \frac{GM' v'}{r^3} \Bigl[ 
\sin\iota\cos\iota\, \delta(\sigma_{\rm ext}-\varpi) 
- \sin\iota\cos\iota\, \delta(\sigma_{\rm ext}+\varpi) \Bigr]. 
\end{align} 
\end{subequations} 
We observe that the delta functions come with a multitude of arguments: the tidal moments possess a large number of frequency components. 

\subsection{Tidal potentials and force densities}     
 
The (Fourier transform of the) gravitoelectric tidal potential is 
\begin{equation} 
\tilde{U}^{\rm tidal} = -\frac{1}{2} \tilde{\cal E}_{jk} x^j x^k, 
\label{U} 
\end{equation} 
where $x^j = (x,y,z)$ is the position vector in the corotating frame, with origin attached to the spheroid's center of mass. It produces a force density 
\begin{equation} 
\tilde{f}^{\rm el}_j = \partial_j \tilde{U}^{\rm tidal}. 
\label{f_el} 
\end{equation} 
The gravitomagnetic tidal potential is \cite{taylor-poisson:08} 
\begin{equation} 
\tilde{U}^{\rm tidal}_j = -\frac{1}{6} \varepsilon_{jkp} \tilde{\cal B}^p_{\ q} x^k x^q, 
\label{Uj}
\end{equation} 
and it produces a force density 
\begin{equation} 
\tilde{f}^{\rm mag}_j = \frac{4}{c^2} \Bigl( -i\sigma_{\rm ext} \tilde{U}^{\rm tidal}_j 
+ \varepsilon_{jkp} \Omega^k \tilde{U}_{\rm tidal} ^p - v^k \partial_j \tilde{U}^{\rm tidal}_k \Bigr), 
\label{fmag} 
\end{equation} 
where $v^j = \varepsilon_{jkp} \Omega^k x^p$ is the unperturbed velocity field. This expression is obtained as follows. 
The force density in the time domain and in the nonrotating frame is (we omit the label ``tidal'' on the vector potential in this calculation) 
\begin{equation} 
f^{\rm mag}_j = \frac{4}{c^2} \Bigl[\partial_t U_j + v^k (\partial_k U_j - \partial_j U_k) \Bigr]; 
\end{equation} 
this is the gravitomagnetic piece of Eq.~(8.119) of Ref.~\cite{poisson-will:14}, with all other post-Newtonian terms discarded. We rewrite the second term with the help of 
\begin{equation} 
\Lie_v U_j = v^k \partial_k U_j + U_k \partial_j v^k = v^k \partial_k U_j - \epsilon_{jkp} \Omega^k U^p. 
\end{equation} 
We interpret the operator $\partial_t + \Lie_v$ acting on $U_j$ as a partial time derivative in the corotating frame. And finally, we perform the Fourier transform to bring the expression to the frequency domain.  

We insert the decompositions of Eq.~(\ref{EB_decomposition}) within the force densities, and use the vector transformation law $f_a = f_j (\partial x^j/\partial x^a)$ to obtain their components in the $(\xi,\nu,\varphi)$ coordinates of Sec.~\ref{sec:coordinates};  because we are now working in the corotating frame, we replace the old azimuthal angle $\phi$ with $\varphi := \phi - \Omega t$. We arrive at expressions of the form 
\begin{equation} 
\tilde{f}^{\rm el}_a = b^2\sum_m \tilde{\cal E}^m\, \gothp_a^m e^{im\varphi}, \qquad 
\tilde{f}^{\rm mag}_a = \frac{b^2\Omega}{c^2} \sum_m \tilde{\cal B}^m\, \gothq_a^m e^{im\varphi}, 
\label{force_m} 
\end{equation} 
where $\gothp_a^m$ and $\gothq_a^m$ are dimensionless functions of $\xi$ and $\nu$; $\gothq_a^m$ also depends linearly upon the dimensionless frequency $w := \sigma_{\rm ext}/\Omega$. Because they are somewhat lengthy, we shall not display here the explicit expressions for  $\gothp_a^m$ and $\gothq_a^m$.  

\subsection{Mode functions and overlap integrals} 

Each inertial mode is described in terms of a Lagrangian displacement vector $\Xi_a = \chi_a e^{-i\sigma t}$. This can be obtained from the hydrodynamic potential $\delta V$ through Eqs.~(\ref{xi_dv}) and (\ref{dv_dV}). Choosing the normalization arbitrarily, we write 
\begin{equation} 
\chi^a = \frac{1}{\kappa^2 - 4} \biggl( g^{ab} - \frac{4}{\kappa^2} e^a e^b 
+ \frac{2i}{\kappa} \varepsilon^{abc} e_c \biggr) \nabla_b \delta V, 
\end{equation} 
where $\kappa := \sigma/\Omega$ is the mode's dimensionless eigenfrequency, $e^a := \Omega^a/\Omega$ a unit vector pointing along the axis of rotation, and 
\begin{equation} 
\delta V = {\sf P}_\ell^m(\xi) {\sf P}_\ell^m(\nu) e^{im\varphi}. 
\end{equation} 
We recall that the modes are labeled with the three integers $\ell$, $m$, and $n$; we omit these labels on $\chi^a$ and $\kappa$ (and other quantities below) to avoid cluttering the notation. 

The overlap integrals of Eq.~(\ref{q_eq}) are 
\begin{equation} 
\tilde{f}^{\rm el} := \int \rho\, \bar{\chi}^a \tilde{f}^{\rm el}_a\, d{\cal V}, \qquad 
\tilde{f}^{\rm mag} := \int \rho\, \bar{\chi}^a \tilde{f}^{\rm mag}_a\, d{\cal V},
\label{f_overlaps} 
\end{equation}
and we express them as 
\begin{equation} 
\tilde{f}^{\rm el} = \rho b^3\, \tilde{\cal E}^m\, \gothh^{\rm el}, \qquad 
\tilde{f}^{\rm mag} = \frac{\rho \Omega b^4}{c^2}\, \tilde{\cal B}^m\, \gothh^{\rm mag}, 
\end{equation} 
in terms of the dimensionless quantities $\gothh^{\rm el}$ and $\gothh^{\rm mag}$. 

The mode norms are 
\begin{equation} 
N = \int \rho\, \bar{\chi}^a \bigl( \chi_a + i\kappa^{-1} \varepsilon_{abc} e^b \chi^c \bigr)\, d{\cal V}, 
\end{equation} 
and we express them as $N = \rho b\, \gothN$, with $\gothN$ dimensionless. The normalized overlap integrals are then 
\begin{equation} 
\tilde{f}^{\rm el}_{\rm norm} := \frac{\tilde{f}^{\rm el}}{\sqrt{N}} 
= \rho^{1/2} b^{5/2}\, \tilde{\cal E}^m\, \frac{\gothh^{\rm el}}{\sqrt{\gothN}} 
\end{equation} 
and
\begin{equation} 
\tilde{f}^{\rm mag}_{\rm norm} := \frac{\tilde{f}^{\rm mag}}{\sqrt{N}} 
= \frac{\rho^{1/2} b^{7/2} \Omega}{c^2}\, \tilde{\cal B}^m\, \frac{\gothh^{\rm mag}}{\sqrt{\gothN}}. 
\end{equation} 
In the cases considered below, for which the overlap integrals do not vanish, the mode norms $\gothN$ are all positive. 

To put the overlap integrals in their final form, we relate the radius parameter $b$ of the $(\xi,\nu,\varphi)$ coordinates to the spheroid's mean radius $R$, by combining Eqs.~(\ref{mean_radius}) and  (\ref{b_def}). We have 
\begin{equation} 
b/R = \frac{\sqrt{4(1+\zeta_0^2) - \kappa^2}}{\sqrt{4-\kappa^2}\, \zeta_0^{1/3} (1+\zeta_0^2)^{1/3}},  
\end{equation} 
and we arrive at 
\begin{equation} 
\tilde{f}^{\rm el}_{\rm norm} = \rho^{1/2} R^{5/2}\, \tilde{\cal E}^m\, \gotho^{\rm el}, \qquad 
\tilde{f}^{\rm mag}_{\rm norm} = \frac{\rho^{1/2} R^{7/2} \Omega}{c^2}\, \tilde{\cal B}^m\, \gotho^{\rm mag} 
\label{f_norm} 
\end{equation} 
with   
\begin{equation} 
\gotho^{\rm el} := (b/R)^{5/2}\, \frac{\gothh^{\rm el}}{\sqrt{\gothN}}, \qquad 
\gotho^{\rm mag} := (b/R)^{7/2}\, \frac{\gothh^{\rm mag}}{\sqrt{\gothN}}. 
\end{equation} 
Because $\gotho^{\rm mag}$ is linear in $w := \sigma_{\rm ext}/\Omega$, we shall express it as $\gotho^{\rm mag}  = \gotho^{\rm mag}_0 + w \gotho^{\rm mag}_1$, with $\gotho^{\rm mag}_0$ and $\gotho^{\rm mag}_1$ independent of $w$.  

\section{Nonvanishing overlap integrals} 
\label{sec:results} 

In Sec.~\ref{sec:vanishing} we shall prove that all overlap integrals between inertial modes and tidal force densities vanish, except for the $r$-mode with $(\ell = 2, m = 1)$ in the case of the gravitoelectric field, and all $\ell = 3$ modes in the case of the gravitomagnetic field. Here we present the exceptional, nonvanishing results for the overlap integrals. 

\begin{table} 
\caption{\label{tab:L2M1} Overlap integrals $\gotho^{\rm el}$ with the gravitoelectric tidal force density for the $(\ell =2, m = 1)$ $r$-mode. The eigenfrequency is constant on the sequence, with $\kappa = \sigma/\Omega = -1$.}
\begin{ruledtabular} 
\begin{tabular}{ccc}
$\zeta_0$ & $\Omega/\sqrt{\pi G \rho}$ & $\gotho^{\rm el}$ \\ 
\hline 
30.0000 & 0.0243316 & $-5.55042\e{-4}$ \\
13.5556 & 0.0537491 & $-2.70876\e{-3}$ \\
8.07407 & 0.0898606 & $-7.57312\e{-3}$ \\
5.33333 & 0.134911 &  $-1.70792\e{-2}$ \\
3.68889 & 0.192007 & $-3.46350\e{-2}$ \\
2.59259 & 0.265222 & $-6.62682\e{-2}$ \\
1.80952 & 0.358983 & $-1.22289\e{-1}$ \\
1.22222 & 0.474603 & $-2.18373\e{-1}$ \\
0.765432 & 0.598303 & $-3.72198\e{-1}$ \\ 
0.400000 & 0.670301 & $-5.99655\e{-1}$ 
\end{tabular} 
\end{ruledtabular} 
\end{table} 

\begin{figure} 
\includegraphics[width=0.4\linewidth]{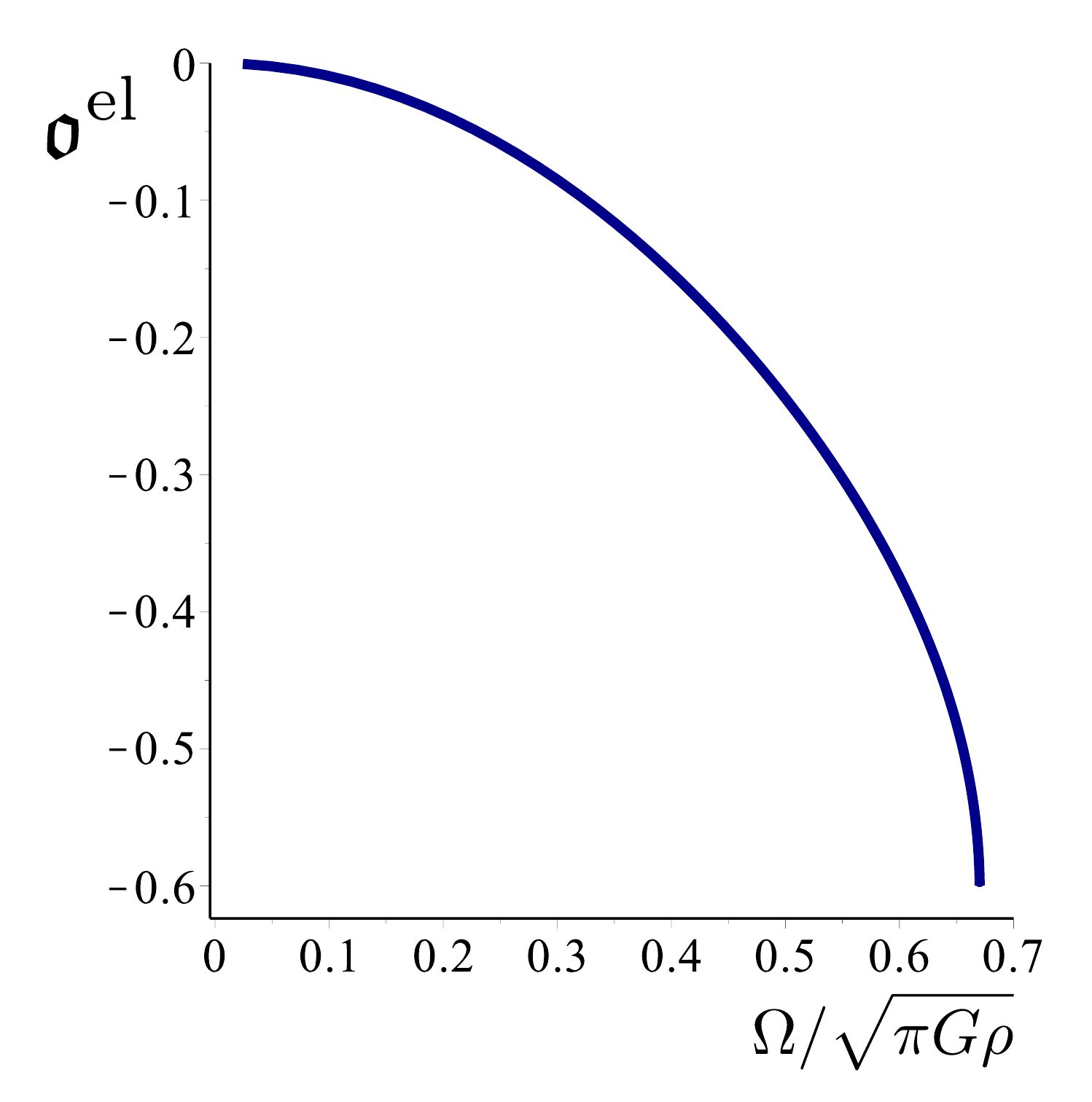}
\caption{Overlap integral $\gotho^{\rm el}$ with the gravitoelectric force density for the $(\ell=2, m=1)$ $r$-mode with eigenfrequency $\kappa = \sigma/\Omega = -1$, plotted as a function of $\Omega/\sqrt{\pi G \rho}$.} 
\label{fig:L2M1} 
\end{figure} 

We begin with the case of the gravitoelectric tidal force. We recall that there is no $\ell = 2$ mode with $m = 2$, and we exclude from our considerations the zero-frequency mode with $m = 0$ (see Appendix~\ref{sec:axi_zerof} for a discussion of this case). The remaining mode with $m = 1$ is the $r$-mode described in some detail in Sec.~\ref{subsec:zerof}; it comes with the eigenfrequency $\kappa = \sigma/\Omega = -1$ for any value of $\zeta_0$. For this mode we obtain 
\begin{subequations} 
\begin{align} 
\gothh^{\rm el} &= -\frac{27}{5} \sqrt{30\pi} \frac{\zeta_0(1 + \zeta_0^2)}{(3 + 4\zeta_0^2)^{5/2}}, \\
\gothN &= \frac{648}{5} \sqrt{3\pi} \frac{\zeta_0(1 + \zeta_0^2)^2}{(3 + 4\zeta_0^2)^{5/2}}, 
\end{align} 
\end{subequations} 
so that 
\begin{equation} 
\gotho^{\rm el} = -\frac{1}{2} \zeta_0^{-1/3} (1 + \zeta_0^2)^{-5/6}. 
\label{overlap_L2} 
\end{equation} 
In the large $\zeta_0$ limit we have that $\gotho^{\rm el} = -\frac{1}{2} \zeta_0^{-2} + O(\zeta_0^{-4})$, or 
\begin{equation} 
\gotho^{\rm el}  \sim -\frac{15}{16} \frac{\Omega^2}{\pi G \rho} \qquad (\mbox{small $\Omega$}). 
\label{overlap_L2_small} 
\end{equation} 
As expected, the overlap integral vanishes in the no-rotation limit. A listing of values is provided in Table~\ref{tab:L2M1}, and a plot of $\gotho^{\rm el}$ as a function of $\Omega/\sqrt{\pi G \rho}$ is shown in Fig.~\ref{fig:L2M1}. 

\begin{table} 
\caption{\label{tab:L3M0} Eigenfrequencies $\kappa = \sigma/\Omega$ of the $(\ell=3, m=0)$ mode with positive frequency, and overlap integrals $\gotho^{\rm mag}$ with the gravitomagnetic tidal force density. In the fourth column, $w = \sigma_{\rm ext}/\Omega$ is the ratio of the external frequency $\sigma_{\rm ext}$ (the frequency of the tidal field in the corotating frame) to the spheroid's angular velocity $\Omega$.}
\begin{ruledtabular} 
\begin{tabular}{cccc}
$\zeta_0$ & $\Omega/\sqrt{\pi G \rho}$ & $\kappa$ & $\gotho^{\rm mag}$ \\ 
\hline 
30.0000 & 0.0243316 & 0.894639 & $-3.08486\e{-1}w - 1.71684\e{-4}$ \\
13.5556 & 0.0537491 & 0.895463 & $-3.08015\e{-1}w - 8.39639\e{-4}$ \\
8.07407 & 0.0898606 & 0.897332 & $-3.06950\e{-1}w - 2.35871\e{-3}$ \\
5.33333 & 0.134911 & 0.901020 & $-3.04856\e{-1}w - 5.36972\e{-3}$ \\
3.68889 & 0.192007 & 0.907955 & $-3.00944\e{-1}w - 1.10824\e{-2}$ \\
2.59259 & 0.265222 & 0.920877 & $-2.93751\e{-1}w - 2.19018\e{-2}$ \\
1.80952 & 0.358983 & 0.945194 & $-2.80544\e{-1}w - 4.29020\e{-2}$ \\
1.22222 & 0.474603 & 0.991661 & $-2.56461\e{-1}w - 8.55320\e{-2}$ \\
0.765432 & 0.598303 & 1.08056 & $-2.14175\e{-1}w -1.78164\e{-1}$ \\ 
0.400000 & 0.670301 & 1.24311 &  $-1.46152\e{-1}w - 4.09681\e{-1}$ 
\end{tabular} 
\end{ruledtabular} 
\end{table} 

\begin{figure} 
\includegraphics[width=0.4\linewidth]{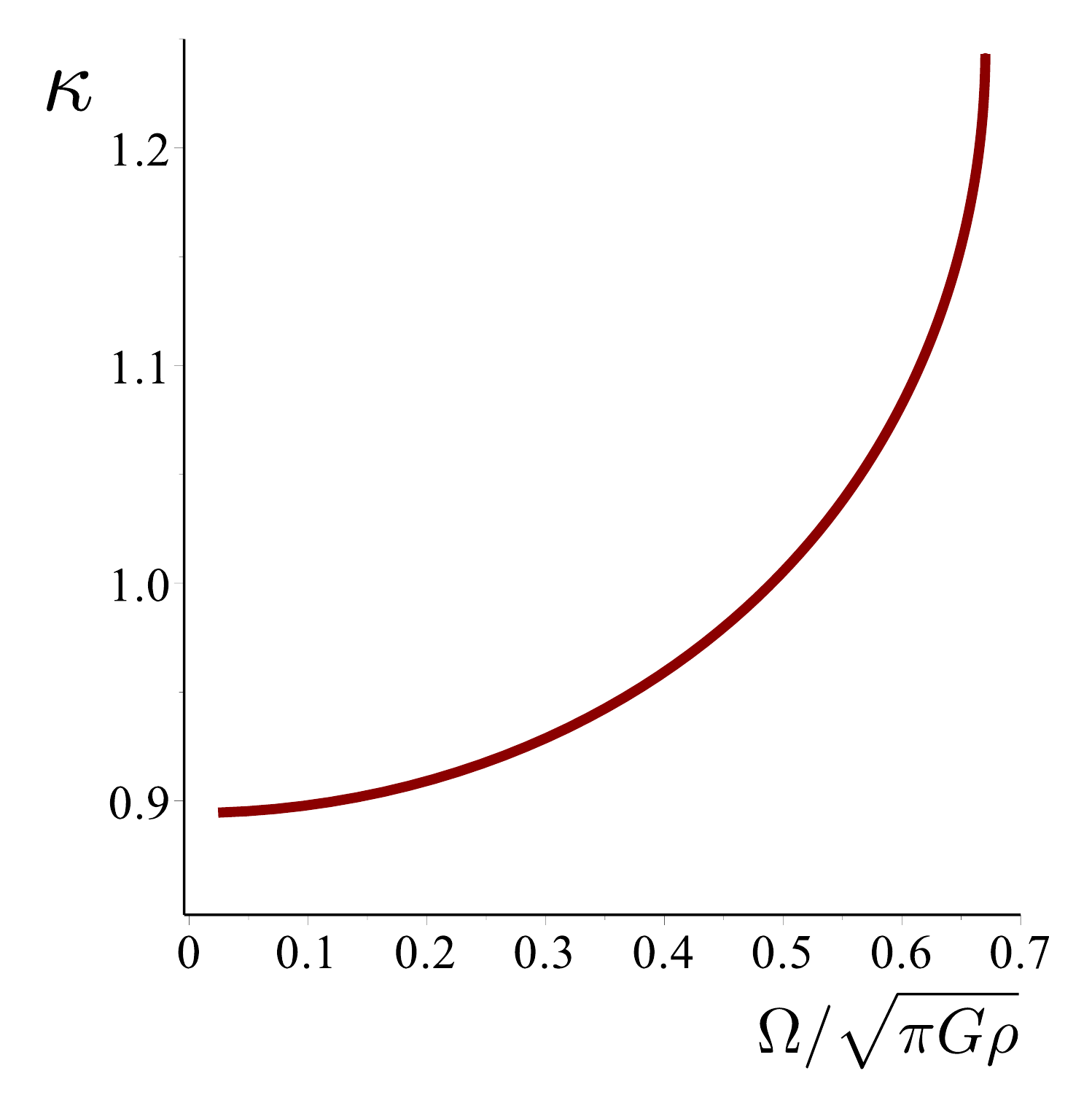}
\includegraphics[width=0.4\linewidth]{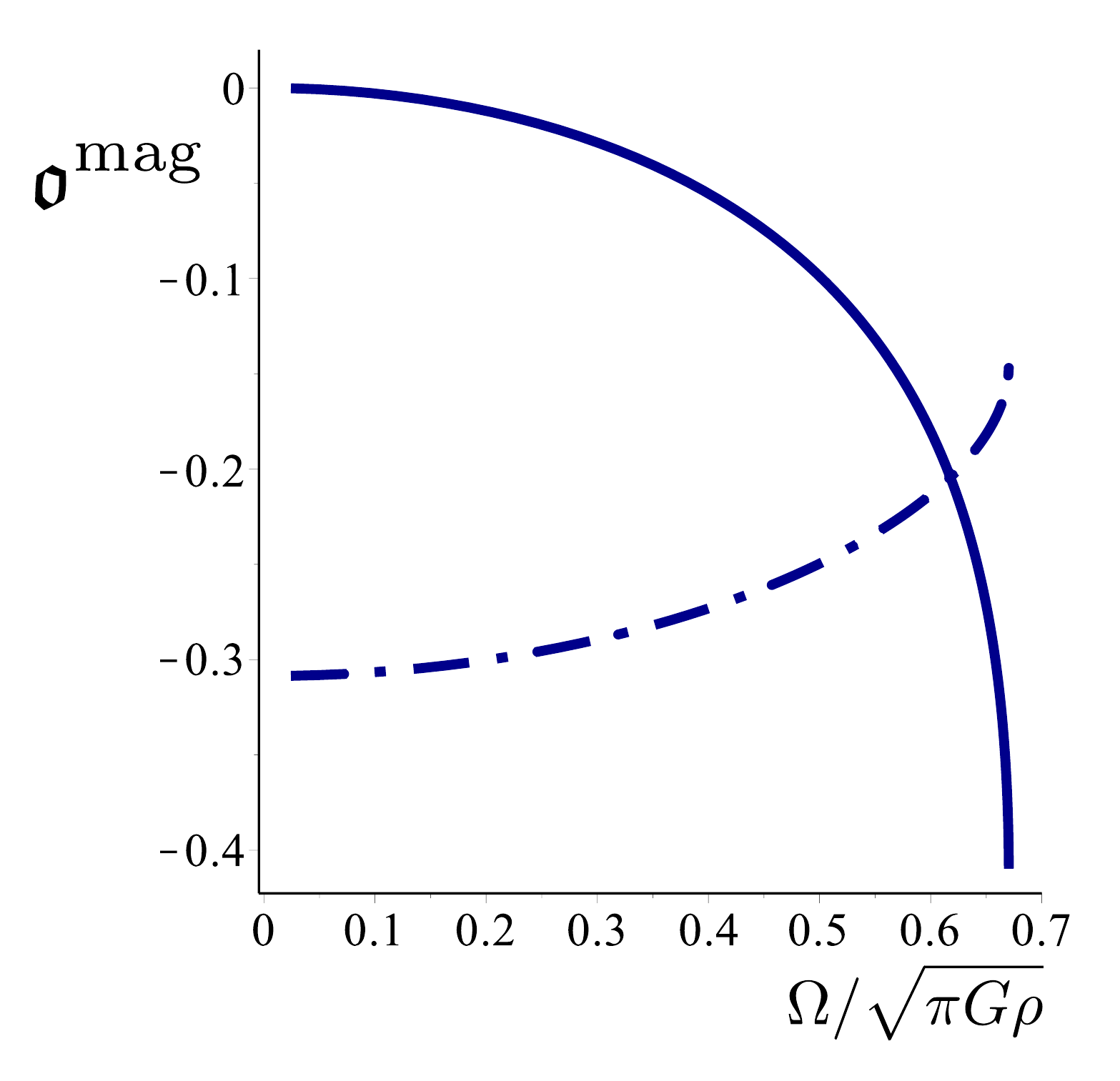}
\caption{Eigenfrequency and overlap integral with the gravitomagnetic force density for the $(\ell=3, m=0)$ mode with positive frequency.  Left: Eigenfrequency $\kappa = \sigma/\Omega$ as a function of $\Omega/\sqrt{\pi G \rho}$. Right: The solid curve shows $\gotho^{\rm mag}_0$, and the dot-dashed curve shows $\gotho^{\rm mag}_1$, such that $\gotho^{\rm mag} = \gotho^{\rm mag}_0 + w\gotho^{\rm mag}_1$, with $w = \sigma_{\rm ext}/\Omega$; both are plotted as functions of $\Omega/\sqrt{\pi G \rho}$.} 
\label{fig:L3M0} 
\end{figure} 

Next we move on to the case of the gravitomagnetic tidal force. We have four modes to examine. The first is actually the pair of modes with $\ell = 3$ and $m = 0$, which comes with positive and negative frequencies of equal magnitude. The mode functions are the same for each mode, and these properties imply that 
\begin{equation} 
(\gotho^{\rm mag}_-)_{\ell = 3}^{m = 1}(\sigma_{\rm ext}) = -(\gotho^{\rm mag}_+)_{\ell = 3}^{m = 1}(-\sigma_{\rm ext});
\end{equation} 
the term on the left is the overlap integral for the negative-frequency mode evaluated at external frequency $\sigma_{\rm ext}$, while the term on the right is the overlap integral for the positive-frequency mode evaluated at frequency $-\sigma_{\rm ext}$. A listing of values for the positive-frequency mode is presented in Table~\ref{tab:L3M0}. Plots of $\kappa = \sigma/\Omega$ and $\gotho^{\rm mag}_+$ as functions of $\Omega/\sqrt{\pi G \rho}$ are displayed in Fig.~\ref{fig:L3M0}. 

\begin{table} 
\caption{\label{tab:L3M1+} Eigenfrequencies $\kappa = \sigma/\Omega$ of the $(\ell=3, m=1)$ mode with positive frequency, and overlap integrals $\gotho^{\rm mag}$ with the gravitomagnetic tidal force density; $w = \sigma_{\rm ext}/\Omega$.}
\begin{ruledtabular} 
\begin{tabular}{cccc}
$\zeta_0$ & $\Omega/\sqrt{\pi G \rho}$ & $\kappa$ & $\gotho^{\rm mag}$ \\ 
\hline
30.0000 & 0.0243316 & 0.176567 &  $3.64526\e{-1}w +3.55844\e{-5}$ \\ 
13.5556 & 0.0537491 & 0.176412 &  $3.64497\e{-1}w + 1.73854\e{-4}$ \\ 
8.07407 & 0.0898606 & 0.176057 &  $3.64434\e{-1}w + 4.87273\e{-4}$ \\ 
5.33333 & 0.134911 & 0.175351 &  $3.64317\e{-1}w + 1.10432\e{-3}$ \\ 
3.68889 & 0.192007 & 0.173997 &  $3.64121\e{-1}w + 2.26006\e{-3}$ \\ 
2.59259 & 0.265222 & 0.171393 &  $3.63842\e{-1}w + 4.39771\e{-3}$ \\ 
1.80952 & 0.358983 & 0.166214 &  $3.63601\e{-1}w + 8.37198\e{-3}$ \\ 
1.22222 & 0.474603 & 0.155361 &  $3.64013\e{-1}w + 1.58488\e{-2}$ \\ 
0.765432 & 0.598303 & 0.131147 &  $3.66214\e{-1}w + 3.02053\e{-2}$ \\ 
0.400000 & 0.670301 & 0.0699222 & $3.42510\e{-1}w + 5.71601\e{-2}$ 
\end{tabular} 
\end{ruledtabular} 
\end{table} 

\begin{figure} 
\includegraphics[width=0.4\linewidth]{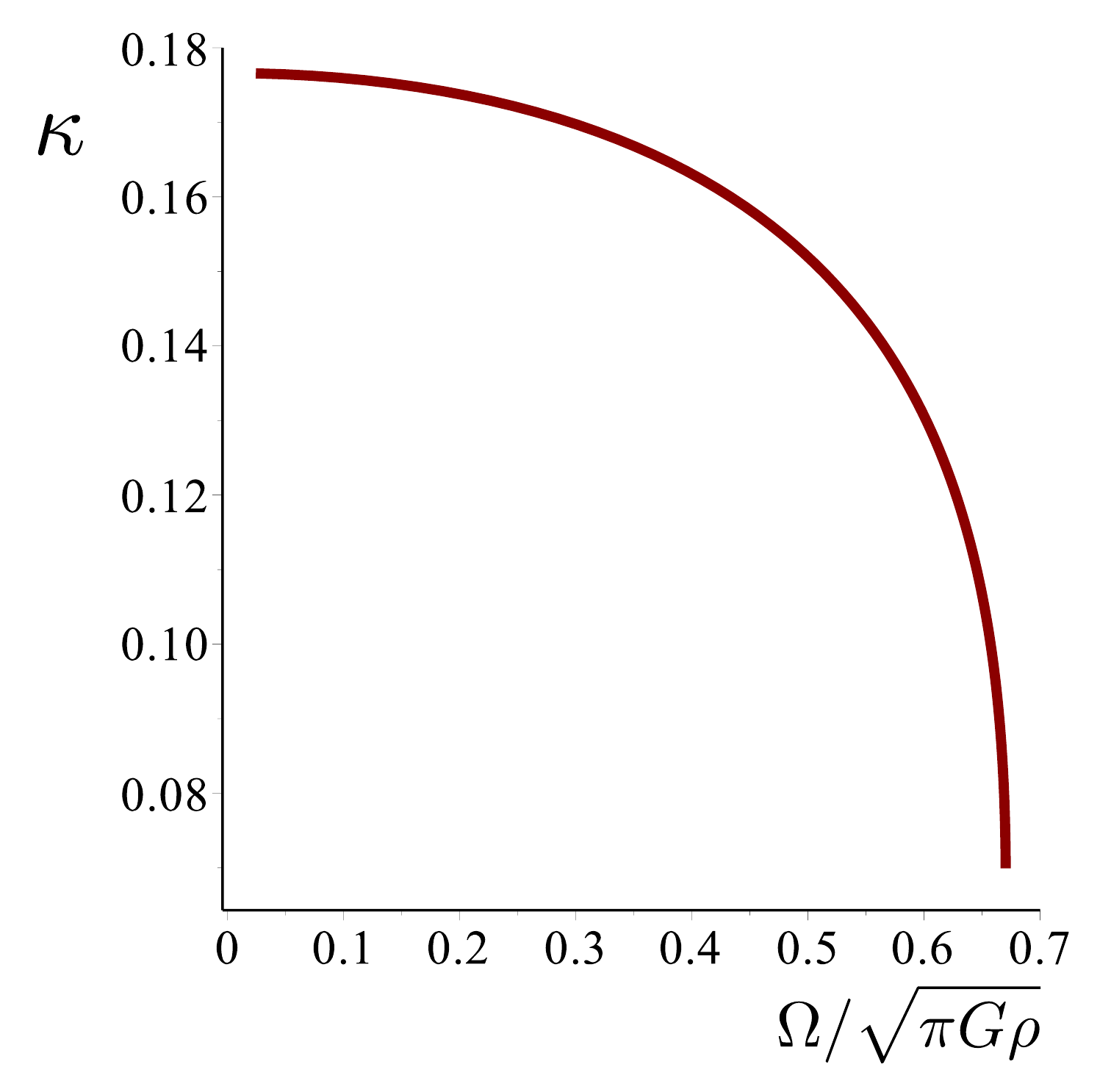}
\includegraphics[width=0.4\linewidth]{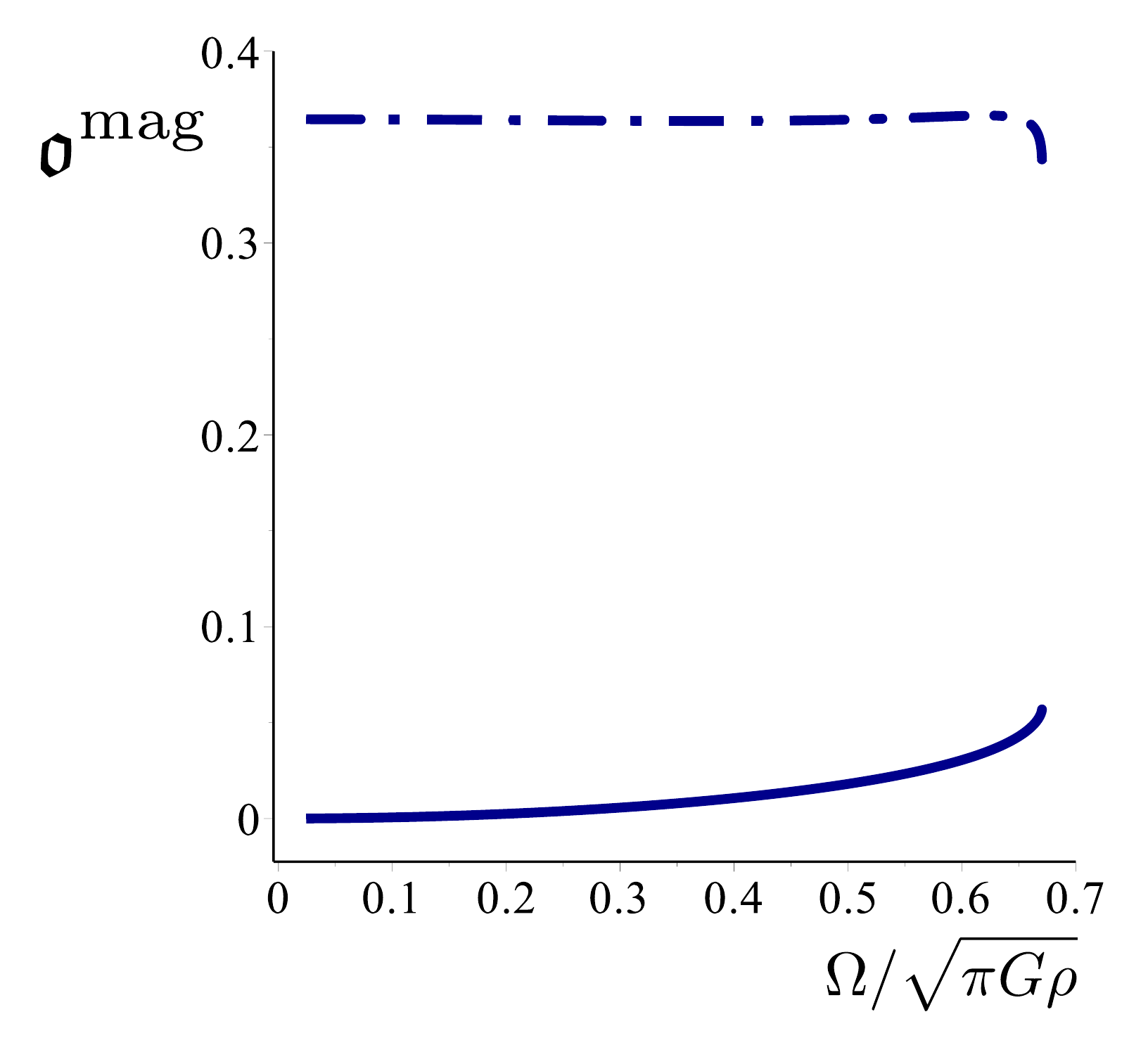}
\caption{Eigenfrequency and overlap integral with the gravitomagnetic force density for the $(\ell=3, m=1)$ mode with positive frequency. Left: Eigenfrequency $\kappa = \sigma/\Omega$ as a function of $\Omega/\sqrt{\pi G \rho}$. Right: The solid curve shows $\gotho^{\rm mag}_0$, and the dot-dashed curve shows $\gotho^{\rm mag}_1$, such that $\gotho^{\rm mag} = \gotho^{\rm mag}_0 + w\gotho^{\rm mag}_1$, with $w = \sigma_{\rm ext}/\Omega$; both are plotted as functions of $\Omega/\sqrt{\pi G \rho}$.} 
\label{fig:L3M1+} 
\end{figure} 

\begin{figure} 
\includegraphics[width=0.4\linewidth]{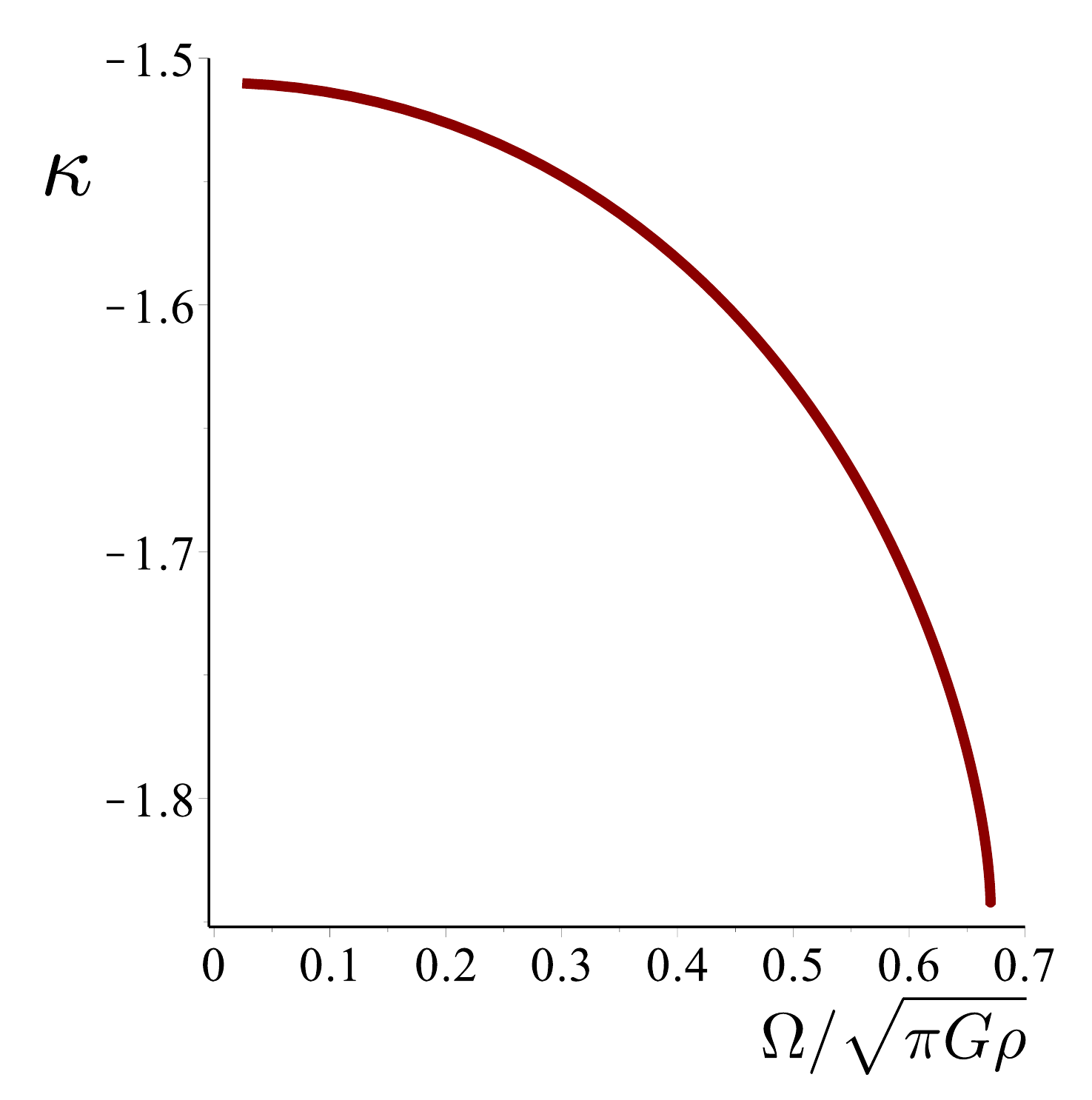}
\includegraphics[width=0.4\linewidth]{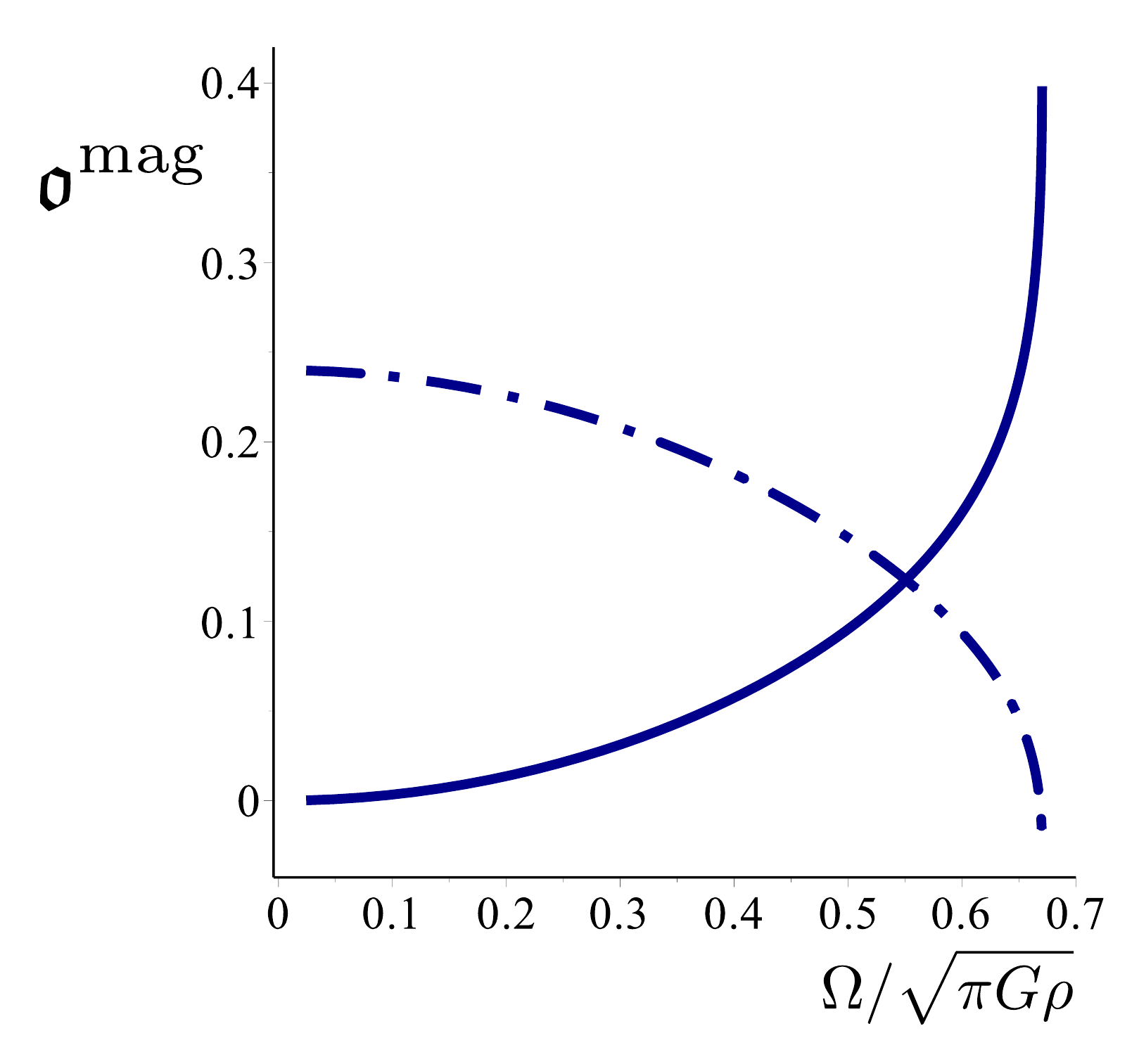}
\caption{Eigenfrequency and overlap integral with the gravitomagnetic force density for the $(\ell=3, m=1)$ mode with negative frequency. Left: Eigenfrequency $\kappa = \sigma/\Omega$ as a function of $\Omega/\sqrt{\pi G \rho}$. Right: The solid curve shows $-\gotho^{\rm mag}_0$, and the dot-dashed curve shows $\gotho^{\rm mag}_1$, such that $\gotho^{\rm mag} = \gotho^{\rm mag}_0 + w\gotho^{\rm mag}_1$, with $w = \sigma_{\rm ext}/\Omega$; both are plotted as functions of $\Omega/\sqrt{\pi G \rho}$.} 
\label{fig:L3M1-} 
\end{figure}

The second and third modes come with $\ell = 3$ and $m = 1$; one has a positive frequency, the other a negative frequency. The mode frequencies and overlap integrals are listed in Tables~\ref{tab:L3M1+} and \ref{tab:L3M1-}. Plots are provided in Figs.~\ref{fig:L3M1+} and \ref{fig:L3M1-}.  

\begin{table} 
\caption{\label{tab:L3M1-} Eigenfrequencies $\kappa = \sigma/\Omega$ of the $(\ell=3, m=1)$ mode with negative frequency, and overlap integrals $\gotho^{\rm mag}$ with the gravitomagnetic tidal force density; $w = \sigma_{\rm ext}/\Omega$.}
\begin{ruledtabular} 
\begin{tabular}{cccc}
$\zeta_0$ & $\Omega/\sqrt{\pi G \rho}$ & $\kappa$ & $\gotho^{\rm mag}$ \\ 
\hline
30.0000 & 0.0243316 & $-1.51017$ & $2.39775\e{-1}w - 2.00253\e{-4}$ \\ 
13.5556 & 0.0537491 & $-1.51108$ & $2.38966\e{-1}w - 9.77700\e{-4}$ \\ 
8.07407 & 0.0898606 & $-1.51312$ & $2.37140\e{-1}w - 2.73606\e{-3}$ \\ 
5.33333 & 0.134911 & $-1.51717$ & $2.33569\e{-1}w - 6.18239\e{-3}$ \\ 
3.68889 & 0.192007 & $-1.52479$ & $2.26966\e{-1}w - 1.25847\e{-2}$ \\  
2.59259 & 0.265222 & $-1.53903$ &  $2.15040\e{-1}w - 2.42609\e{-2}$ \\ 
1.80952 & 0.358983 & $-1.56590$ & $1.93798\e{-1}w - 4.54913\e{-2}$ \\ 
1.22222 & 0.474603 & $-1.61708$ &  $1.56809\e{-1}w - 8.43235\e{-2}$ \\ 
0.765432 & 0.598303 & $-1.71101$ & $9.47456\e{-2}w - 1.59035\e{-1}$ \\  
0.400000 & 0.670301 & $-1.84317$ & $-2.64212\e{-2}w -3.98212\e{-1}$ 
\end{tabular} 
\end{ruledtabular} 
\end{table} 

\begin{figure} 
\includegraphics[width=0.4\linewidth]{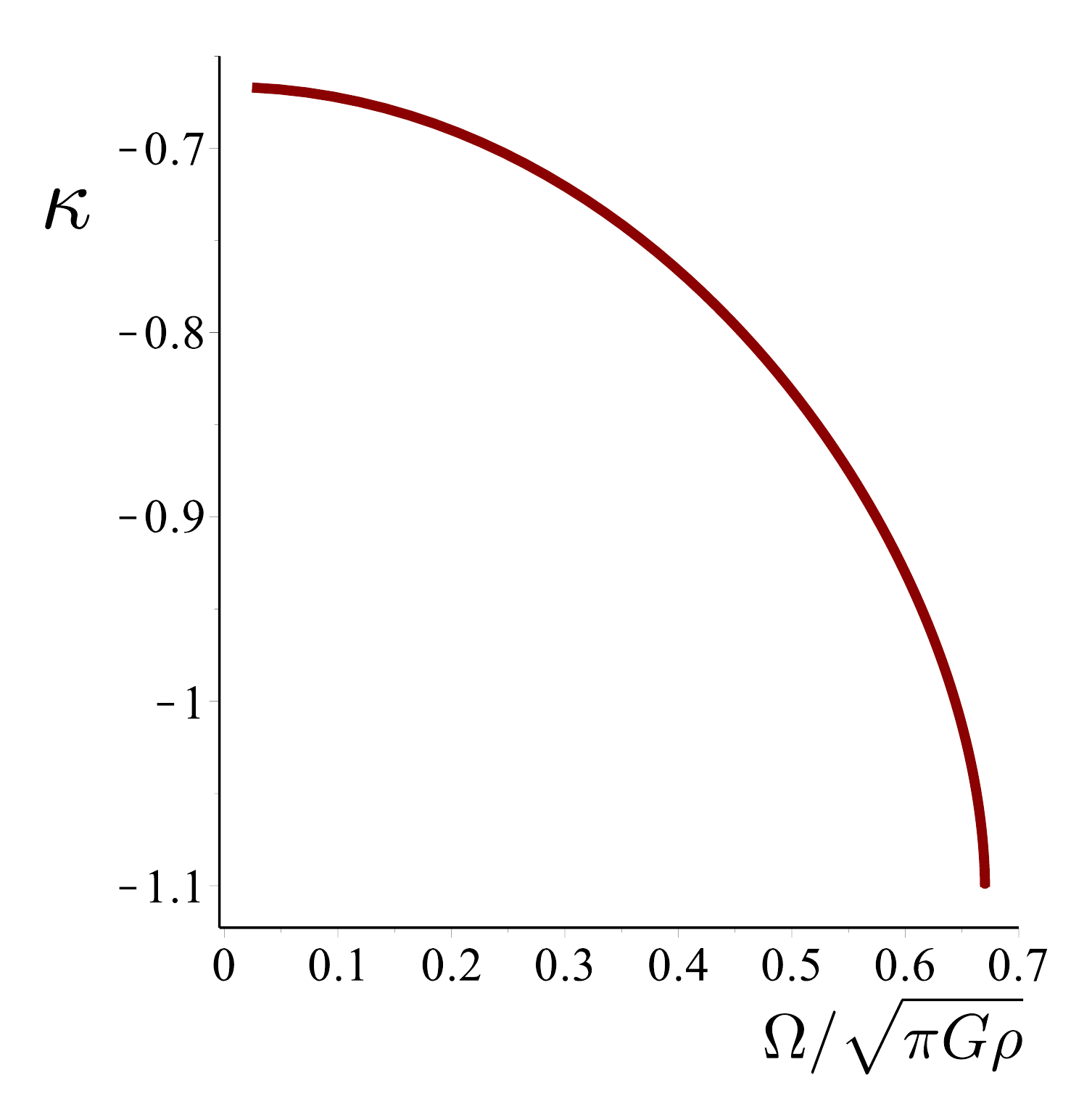}
\includegraphics[width=0.4\linewidth]{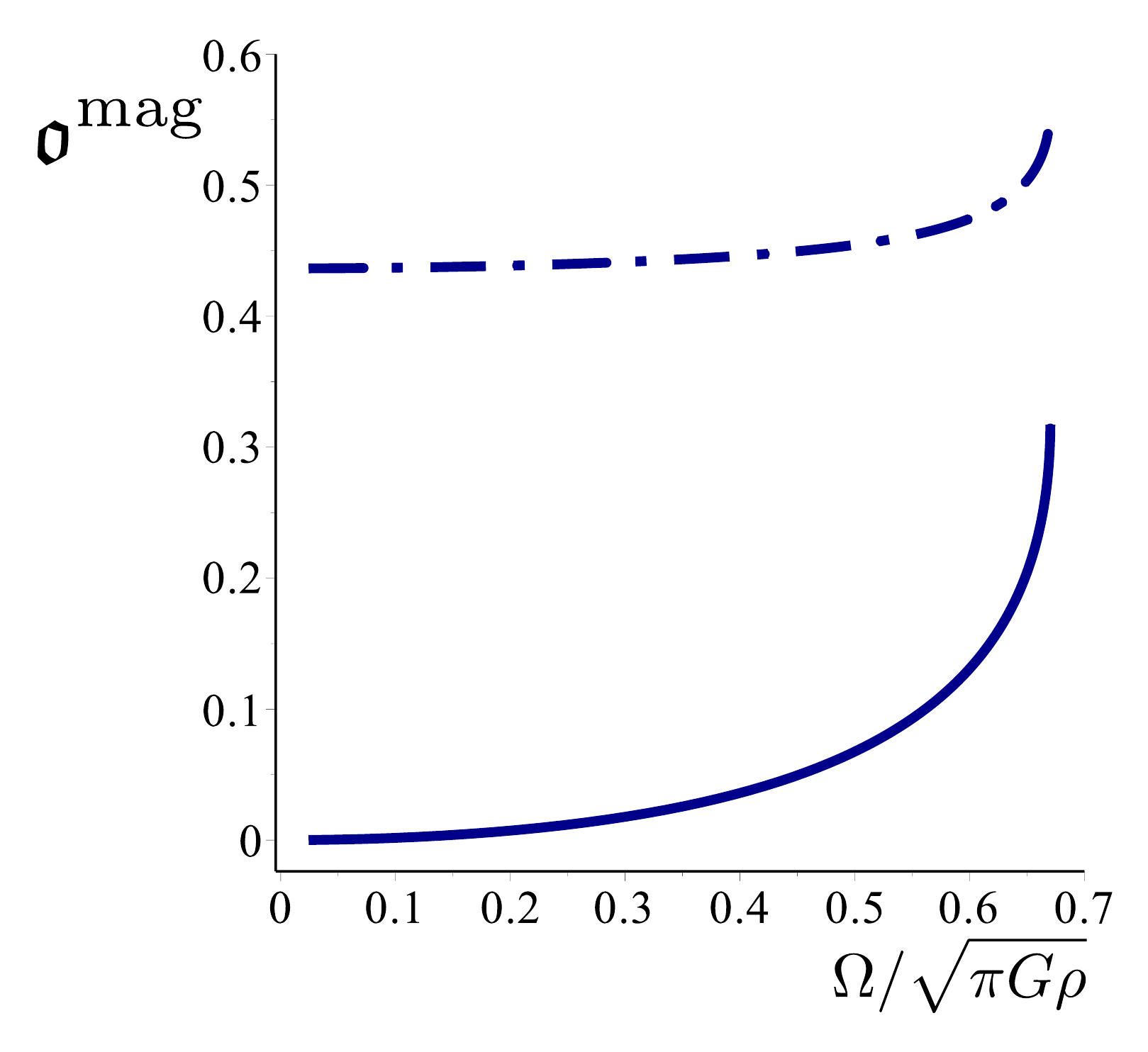}
\caption{Eigenfrequency and overlap integral with the gravitomagnetic force density for the $(\ell=3, m=2)$ $r$-mode. Left: Eigenfrequency $\kappa = \sigma/\Omega$ as a function of $\Omega/\sqrt{\pi G \rho}$. Right: The solid curve shows $-\gotho^{\rm mag}_0$, and the dot-dashed curve shows $\gotho^{\rm mag}_1$, such that $\gotho^{\rm mag} = \gotho^{\rm mag}_0 + w\gotho^{\rm mag}_1$, with $w = \sigma_{\rm ext}/\Omega$; both are plotted as functions of $\Omega/\sqrt{\pi G \rho}$.} 
\label{fig:L3M2} 
\end{figure}

The fourth and final mode is the $r$-mode with $\ell = 3$ and $m=2$. Because this mode has a negative frequency in the 
star's corotating frame ($\kappa < 0$), but a positive frequency in the nonrotating frame ($\kappa + 2 > 0$), it is subject to the Chandrasekhar-Friedman-Schutz instability \cite{chandrasekhar:70, friedman-schutz:78b}: the mode grows as it emits gravitational waves. The frequencies and overlap integrals are tabulated in Table~\ref{tab:L3M2} and plotted in Fig.~\ref{fig:L3M2}.   

\begin{table} 
\caption{\label{tab:L3M2} Eigenfrequencies $\kappa = \sigma/\Omega$ of the $(\ell=3, m=2)$ $r$-mode, and overlap integrals $\gotho^{\rm mag}$ with the gravitomagnetic tidal force density; $w = \sigma_{\rm ext}/\Omega$.}
\begin{ruledtabular} 
\begin{tabular}{cccc}
$\zeta_0$ & $\Omega/\sqrt{\pi G \rho}$ & $\kappa$ & $\gotho^{\rm mag}$ \\ 
\hline
30.0000 & 0.0243316 & $-0.667007$ &  $4.36466\e{-1}w - 1.00581\e{-4}$ \\ 
13.5556 & 0.0537491 & $-0.668327$ & $4.36585\e{-1}w - 4.92688\e{-4}$ \\ 
8.07407 & 0.0898606 & $-0.671319$ & $4.36853\e{-1}w - 1.38904\e{-3}$ \\ 
5.33333 & 0.134911 & $-0.677202$ &  $4.37381\e{-1}w - 3.18435\e{-3}$ \\  
3.68889 & 0.192007 & $-0.688190$ & $4.38372\e{-1}w - 6.65591\e{-3}$ \\ 
2.59259 & 0.265222 & $-0.708396$ & $4.40221\e{-1}w - 1.34489\e{-2}$ \\
1.80952 & 0.358983 & $-0.745445$ & $4.43816\e{-1}w - 2.73401\e{-2}$ \\ 
1.22222 & 0.474603 & $-0.812607$ & $4.51687\e{-1}w - 5.77031\e{-2}$ \\ 
0.765432 & 0.598303 & $-0.928110$ & $4.73774\e{-1}w - 1.29307\e{-1}$ \\  
0.400000 & 0.670301 & $-1.10071$ & $5.59537\e{-1}w - 3.17079\e{-1}$ 
\end{tabular} 
\end{ruledtabular} 
\end{table} 

The tables and figures reveal that with the overlap integrals written as $\gotho^{\rm mag} = \gotho^{\rm mag}_0 + w \gotho^{\rm mag}_1$, where $w := \sigma_{\rm ext}/\Omega$, $\gotho^{\rm mag}_0$ is orders of magnitude smaller than $\gotho^{\rm mag}_1$ when $\zeta_0$ is large ($\Omega$ small). This implies that the overlap integral is dominated by the induction piece of the gravitomagnetic tidal force, given by the $-i\sigma_{\rm ext} \tilde{U}_j^{\rm tidal}$ term in Eq.~(\ref{fmag}). They eventually become comparable as $\zeta_0$ decreases to small values and $\Omega$ becomes large.   

\section{Vanishing overlap integrals} 
\label{sec:vanishing}  

For a generic and unspecified value of $\ell$, the computation of the overlap integrals $\tilde{f}^{\rm el}$ and $\tilde{f}^{\rm mag}$, defined by Eq.~(\ref{f_overlaps}), requires extensive manipulation of associated Legendre functions ${\sf P}_\ell^m(x)$, where $x$ stands for either $\xi$ or $\nu$. In this section we sketch these manipulations, which lead us to conclude that except for the special cases examined previously in Sec.~\ref{sec:results}, all overlap integrals between inertial modes of Maclaurin spheroids and quadrupolar tidal forces vanish. 

The integrands in Eq.~(\ref{f_overlaps}) feature associated Legendre functions and their first derivatives. For $m=0$ and $m=2$ we make use of the recursion relation [NIST Eq.~(14.10.4)] 
\begin{equation} 
(1-x^2) \frac{d {\sf P}_\ell^m}{dx} = -(\ell-m+1) {\sf P}_{\ell+1}^m + (\ell+1) x {\sf P}_\ell^m 
\label{recursion1} 
\end{equation} 
to eliminate the differentiated functions. The factors of $(1-x^2)^{-1}$ generated by the recursion relation cancel out factors of $(1-x^2)$ that occur within the integrands. The end result consists of a sum of basic factors drawn from the set  
\begin{equation} 
{\cal S} = \Bigl\{ {\sf P}_\ell^m, x{\sf P}_\ell^m, x^2 {\sf P}_\ell^m, x^3 {\sf P}_\ell^m, x {\sf P}_{\ell+1}^m,  
x^2 {\sf P}_{\ell+1}^m \Bigl\}. 
\label{set} 
\end{equation} 

For $m=1$ we begin instead by turning the associated functions into polynomials with the help of the definition ${\sf P}_\ell^1 := \sqrt{1-x^2}\, d{\sf P}_\ell/dx$. In this case the square-root factor cancels out similar factors that occur in the denominator of each integrand. The second derivatives of the polynomials are then eliminated with Legendre's differential equation, and the first derivatives are again eliminated with the recursion relation. Once more we end up with a sum of terms from the set $\cal S$. 

The integrals over members of $\cal S$ are carried out with the identities listed in Appendix \ref{sec:legendre}. For $\tilde{f}^{\rm el}$ we arrive at something proportional to 
\begin{equation} 
\frac{1}{\ell-2} \Bigl[ {\sf P}_{\ell+1}^m(-\xi_0) {\sf P}_\ell^m(\xi_0)
+ {\sf P}_{\ell+1}^m(\xi_0) {\sf P}_\ell^m(-\xi_0) \Bigr], 
\end{equation} 
with $m = 0, 2$. This is zero whenever $\ell \neq 2$, by virtue of the identity [NIST Eq.~(14.7.17)] ${\sf P}_\ell^m(-x) = (-1)^{\ell-m} {\sf P}_\ell^m(x)$. The overlap integrals, therefore, vanish except when $\ell  = 2$.  

For $\tilde{f}^{\rm mag}$ we arrive instead at something proportional to 
\begin{equation} 
\frac{1}{\ell-3} \Bigl[ {\sf P}_{\ell+1}^m(-\xi_0) {\sf P}_\ell^m(\xi_0)
+ {\sf P}_{\ell+1}^m(\xi_0) {\sf P}_\ell^m(-\xi_0) \Bigr] 
\end{equation} 
with $m = 0, 2$. This is zero whenever $\ell \neq 3$, and we conclude that the overlap integrals vanish except when $\ell  = 3$.  

Combining these results, we establish that except for modes with $\ell = 2$ in the case of the gravitoelectric force, and modes with $\ell = 3$ in the case of the gravitomagnetic force, all overlap integrals between inertial modes of Maclaurin spheroids and quadrupolar tidal force densities vanish.  

\section{Ratio of gravitoelectric to gravitomagnetic tidal driving} 
\label{sec:ratio} 

In a typical binary system, even a compact one, the Newtonian tidal force acting on a rotating star is much larger than the post-Newtonian, gravitomagnetic force. But it couples more weakly to the star's inertial modes: its overlap integral is suppressed by a factor of $\Omega^2/(\pi  G \rho)$ relative to a naive expectation based on dimensional analysis. In this section we examine the conditions under which the gravitoelectric force dominates the tidal driving of inertial modes of a Maclaurin spheroid. 

Taking Eq.~(\ref{O2}) and inserting relevant scales, we find that the spheroid's rotational frequency $f_{\rm rot} = \Omega/(2\pi)$ is given by 
\begin{equation} 
f_{\rm rot} = 2.66\times 10^3\, \biggl( \frac{M}{1.4\ M_\odot} \biggr)^{1/2} 
\biggl( \frac{10\ \mbox{km}}{R} \biggr)^{3/2}\, \zeta_0^{1/2} 
\bigl[ (1+3\zeta_0^2) \arccot\zeta_0 - 3\zeta_0 \bigr]^{1/2}\ \mbox{Hz}. 
\label{frot} 
\end{equation} 
We shall be interested in a range of rotational frequencies between $50\ \mbox{Hz}$ and $400\ \mbox{Hz}$, the upper bound representing a rather extreme extrapolation of known binary-pulsar frequencies. For this interval, and for the fiducial values adopted for $M$ and $R$ in Eq.~(\ref{frot}),  we have that $\zeta_0$ varies between approximately $27.4$ and $3.30$. We shall take this range of values to be much larger than one, and approximate the overlap integrals with their behavior at large $\zeta_0$. Because the relative errors scale as $\zeta_0^{-2}$, this approximation is perfectly adequate for our purposes here.  

The normalized overlap integrals are given by Eq.~(\ref{f_norm}), in terms of dimensionless amplitudes $\gotho^{\rm el}$ and $\gotho^{\rm mag}$ that were calculated in Sec.~\ref{sec:results}. We take the tidal moments to be those created by a companion body of mass $M'$ on a circular orbit of radius $p$, as displayed in Eqs.~(\ref{E_binary}) and (\ref{B_binary}); the orbital angular velocity $\varpi$ is given by Eq.~(\ref{orbital_frequency}), and the corresponding orbital frequency is $f_{\rm orb} = \varpi/(2\pi)$. 

As measures of the gravitoelectric and gravitomagnetic tidal driving of inertial modes, we adopt
\begin{equation} 
\F^{\rm el} := \rho^{1/2} R^{5/2} \sum_{\rm modes} \sum_{\sigma_{\rm ext}} 
\mbox{max}(\iota) \bigl| \gotho^{\rm el} \tilde{\cal E}^m \bigr|, \qquad 
\F^{\rm mag} := \frac{\rho^{1/2} R^{7/2} \Omega}{c^2} \sum_{\rm modes} \sum_{\sigma_{\rm ext}} 
\mbox{max}(\iota) \bigl| \gotho^{\rm el} \tilde{\cal B}^m \bigr|. 
\label{Fmeasures} 
\end{equation} 
In words, we take the normalized overlap integral for each contributing frequency and each relevant mode, maximize its (complex) absolute value with respect to the inclination angle, and sum over all contributions. As a measure of the ratio between gravitoelectric and gravitomagnetic tidal driving, we take 
\begin{equation} 
\R := \frac{\F^{\rm el}}{\F^{\rm mag}}. 
\label{Ratio} 
\end{equation} 
In these manipulations we choose to account only for modes with $m \geq 0$, and only for one member of the complex-conjugate pair of modes with $m=0$. It is also understood that the delta functions attached to $\tilde{\cal E}^m$ and $\tilde{\cal B}^m$ are discarded in Eq.~(\ref{Fmeasures}); we achieve this formally by integrating over $\sigma_{\rm ext}$. 

For $\F^{\rm el}$ there is only one contributing mode, the $r$-mode with $\ell = 2$ and $m = 1$. The dimensionless overlap integral is given by Eq.~(\ref{overlap_L2_small}) in the large-$\zeta_0$ approximation. After maximizing each term in Eq.~(\ref{E_binary}) with respect to $\iota$, we arrive at 
\begin{equation} 
\F^{\rm el} = \frac{1}{8} \sqrt{30\pi} \biggl( 1 + \frac{3}{2} \sqrt{3} \biggr) \rho^{1/2} R^{11/2} 
\frac{M' \Omega^2}{M r^3}. 
\end{equation} 
For $\F^{\rm mag}$ there are four contributing modes, those with $\ell = 3$ and $m = \{0, 1 ,2\}$; there are two distinct modes with $m = 1$. Taking all of them into account, and all the frequency components of the gravitomagnetic tidal moments of Eq.~(\ref{B_binary}), we obtain 
\begin{equation} 
\F^{\rm mag} = \frac{1}{20} \sqrt{30\pi} \rho^{1/2} R^{7/2} \frac{GM' \Omega v'}{c^2 r^3} \Gamma, 
\end{equation} 
where $v' = r\varpi$ is the orbital velocity, and 
\begin{align} 
\Gamma &:= 2\sqrt{6} \bigl[ |\gotho^{\rm mag}(\sigma_{\rm ext}=\varpi)|^{m=0} 
+ |\gotho^{\rm mag}(\sigma_{\rm ext}=-\varpi)|^{m=0} \bigr] 
\nonumber \\ & \quad \mbox{} 
+ 8 \bigl[ |\gotho^{\rm mag}(\sigma_{\rm ext}=\varpi-\Omega)|^{m=1}_+    
+ |\gotho^{\rm mag}(\sigma_{\rm ext}=-\varpi-\Omega)|^{m=1}_+ \bigr] 
\nonumber \\ & \quad \mbox{} 
+ 8 \bigl[ |\gotho^{\rm mag}(\sigma_{\rm ext}=\varpi-\Omega)|^{m=1}_-    
+ |\gotho^{\rm mag}(\sigma_{\rm ext}=-\varpi-\Omega)|^{m=1}_-\bigr] 
\nonumber \\ & \quad \mbox{} 
+ 3\sqrt{3} \bigl[ |\gotho^{\rm mag}(\sigma_{\rm ext}=\varpi-2\Omega)|^{m=2} 
+ |\gotho^{\rm mag}(\sigma_{\rm ext}=-\varpi-2\Omega)|^{m=2} \bigr]. 
\label{Gamma_def}
\end{align} 
In $\Gamma$, the first set of terms comes from one member of the $m=0$ pair of modes, the second set comes from the positive-frequency $m=1$ mode, the third comes from the negative-frequency $m=1$ mode, and the fourth set of terms comes from the $m=2$ $r$-mode. We recall that $\gotho^{\rm mag}$ depends linearly upon $w := \sigma_{\rm ext}/\Omega$; in the large-$\zeta_0$ approximation we take it to be equal to the first entry in Tables~\ref{tab:L3M0}, \ref{tab:L3M1+}, \ref{tab:L3M1-} and \ref{tab:L3M2}. 

The ratio of Eq.~(\ref{Ratio}) is 
\begin{equation} 
\R = \frac{5}{2} \biggl( 1 + \frac{3}{2} \sqrt{3} \biggr)  \frac{1}{\Gamma} 
\frac{c^2 \Omega R^2}{GM v'}. 
\end{equation} 
This is a re-statement of Eq.~(\ref{ratio_intro}). As was pointed out in Sec.~\ref{sec:intro}, the overall scaling with $c^2\Omega R^2/(GM v')$ can be deduced from a simple dimensional analysis that accounts for the vanishing of $\tilde{f}^{\rm el}$ in the small-$\Omega$ approximation. The numerical factors are contributed by the full calculation presented here. 

\begin{figure} 
\includegraphics[width=0.4\linewidth]{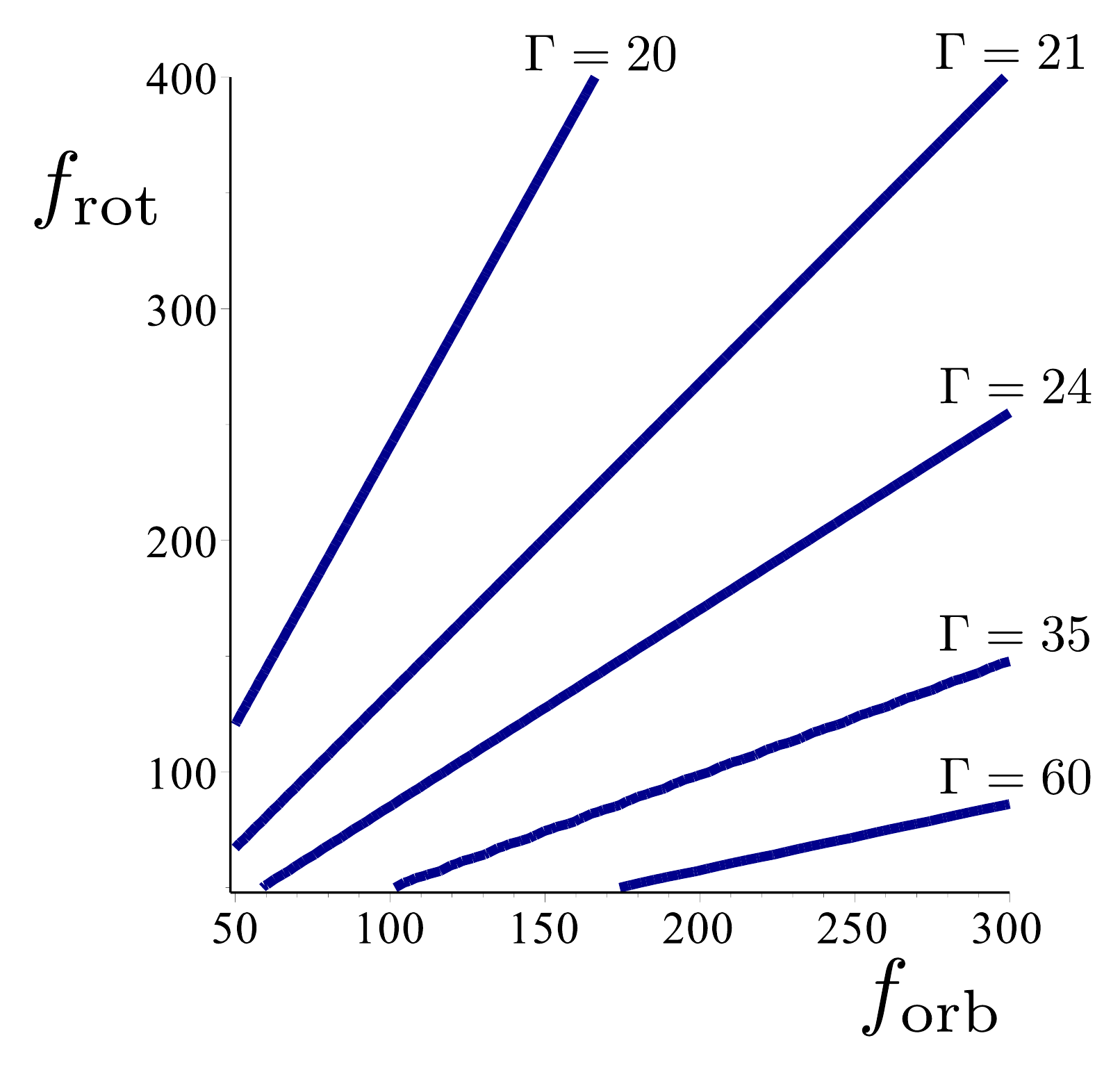}
\caption{Prefactor $\Gamma$ in the reduced ratio $\SSS$ of gravitoelectric to gravitomagnetic tidal driving. In descending order from the highest to lowest curve, we have contours of $\Gamma = \{ 20, 21, 24, 35, 60 \}$, respectively.} 
\label{fig:Gamma} 
\end{figure} 

In Sec.~\ref{sec:intro} we also expressed $\R$ as 
\begin{equation} 
\R = \biggl( \frac{R}{10\ \mbox{km}} \biggr)^2 \biggl( \frac{1.4\ M_\odot}{M} \biggr) 
\biggl( \frac{2.8\ M_\odot}{M_{\rm tot}} \biggr)^{1/3}\, \SSS 
\end{equation} 
with 
\begin{equation} 
\SSS := 0.178 \biggl( \frac{25}{\Gamma} \biggr) \biggl( \frac{f_{\rm rot}}{100\ \mbox{Hz}} \biggr) 
\biggl( \frac{100\ \mbox{Hz}}{f_{\rm orb}} \biggr)^{1/3}. 
\label{Sdef} 
\end{equation} 
We pointed out that the prefactor of $\Gamma^{-1}$ contributes significantly to the variation of $\SSS$ over the frequency intervals considered here. We illustrate this with the contour plot of Fig.~\ref{fig:Gamma}.  A contour plot of $\SSS$ was presented in Fig.~\ref{fig:Ratio}, along with a discussion of its significance. 

\begin{acknowledgments} 
We are grateful to Eanna Flanagan, who helped us understand the meaning of the $\ell=2$, $m=1$ $r$-mode, in view of its zero frequency in the nonrotating frame, as well as the physical significance of Jordan-chain modes (encountered in Appendix \ref{sec:axi_zerof}). This work was supported by the Natural Sciences and Engineering Research Council of Canada.  
\end{acknowledgments} 

\appendix 

\section{Frequently occurring symbols} 
\label{sec:symbols} 

For the convenience of the reader, we list in Table~\ref{tab:symbols} the symbols that occur most frequently within the paper, and provide a brief description.  

\begin{table} 
\caption{\label{tab:symbols} Frequently occurring symbols, grouped according to various themes.} 
\begin{ruledtabular} 
\begin{tabular}{ll}
Symbol & Description \\ 
\hline
\multicolumn{2}{c}{Maclaurin spheroid} \\
$M$ & Mass \\  
$R$ & Mean radius, Eq.~(\ref{mean_radius}) \\  
$\rho$ & Mass density \\
$\Omega$ & Angular velocity \\ 
$f_{\rm rot} = \Omega/(2\pi)$ & Rotational frequency \\ 
$a$ & Focal radius, Eq~(\ref{Re_Rp}) \\ 
$\zeta_0$ & Ellipticity parameter \\ 
$h$ & Specific enthalpy \\ 
& \\ 
\multicolumn{2}{c}{Companion star and orbit} \\
$M'$ & Mass \\ 
$M_{\rm tot} = M + M'$ & Total mass \\ 
$r$ & Orbital radius \\ 
$\varpi$ & Orbital angular velocity, Eq.~(\ref{orbital_frequency}) \\ 
$f_{\rm orb} = \varpi/(2\pi)$ & Orbital frequency \\ 
$v' = r \varpi$ & Orbital velocity \\ 
& \\ 
\multicolumn{2}{c}{Fluid perturbations} \\
$\Xi^a = \chi^a(x^1, x^2)\, e^{im\phi} e^{-i\omega t}$ & Lagrangian displacement \\ 
$\delta v^a$ & Velocity perturbation \\ 
$\delta V = \delta (h - U)$ & Perturbation in hydrodynamic potential \\ 
$\delta U$ & Perturbation in Newtonian potential \\ 
$\omega$ & Mode frequency in nonrotating frame \\ 
$\sigma = \omega - m\Omega$ & Mode frequency in corotating frame \\ 
$\kappa = \sigma/\Omega$ & Dimensionless mode frequency \\ 
$\sigma_{\rm ext}$ & Frequency of tidal field in corotating frame \\ 
& \\ 
\multicolumn{2}{c}{Coordinate systems} \\
$(\zeta, \mu, \phi)$ & Defined by Eq.~(\ref{coords_2}) \\ 
$(\xi, \nu, \phi)$ & Defined by Eq.~(\ref{coords_3}) \\ 
$b$ & Radius parameter, Eq.~(\ref{b_def}) \\ 
$\xi_0$ & Surface value of $\xi$, Eq.~(\ref{xi0_def}) \\ 
$\varphi = \phi - \Omega t$ & Azimuthal angle in corotating frame \\ 
& \\ 
\multicolumn{2}{c}{Gravitational potentials and tidal moments} \\
$U$ & Newtonian (gravitoelectric) potential \\ 
$U^j$ & Post-Newtonian (gravitomagnetic) potential \\ 
${\cal E}_{jk}$ & Gravitoelectric tidal tensor, Eqs.~(\ref{E_binary}), (\ref{U}) \\ 
${\cal B}_{jk}$ & Gravitomagnetic tidal tensor, Eqs.~(\ref{B_binary}), (\ref{Uj}) \\ 
& \\ 
\multicolumn{2}{c}{Associated Legendre functions} \\
${\sf P}_\ell^m(x)$ & When $-1 \leq x \leq 1$ \\ 
$P_\ell^m(z)$ & When $z \geq 1$ or is complex 
\end{tabular} 
\end{ruledtabular} 
\end{table} 

\section{Axisymmetric modes with zero frequency}
\label{sec:axi_zerof} 

The two-potential formalism of Sec.~\ref{subsec:2pot} breaks down when $\sigma = \kappa\Omega$ vanishes, because Eq.~(\ref{dEuler2}) can no longer be inverted. Simultaneously, the $(\xi,\nu,\phi)$ coordinates of Sec.~\ref{sec:coordinates} are not defined when $\kappa = 0$. The case of axisymmetric modes ($m=0$) with $\kappa = 0$ therefore requires a separate treatment, which we provide in this Appendix. 

\subsection{Mode equations} 

We require all perturbation variables $(\delta v^a, \delta\rho, \delta h, \delta U)$ to be axisymmetric (invariant under rotations about the $z$-axis) and time-independent. To satisfy Eq.~(\ref{xi_vs_dv}), however, the Lagrangian displacement vector must be given by 
\begin{equation} 
\Xi^a = (\delta v^a) t + \eta^a, 
\end{equation} 
where $\eta^a$ is independent of time. Substitution within Eq.~(\ref{drho}) yields 
\begin{equation} 
\nabla_a (\rho \delta v^a) = 0, \qquad  \delta \rho = -\nabla_a (\rho \eta^a). 
\end{equation} 
Applying these to a Maclaurin spheroid by invoking Eq.~(\ref{EOS}), we obtain the interior equations 
\begin{equation} 
\nabla_a \delta v^a = 0, \qquad \nabla_a \eta^a = 0 
\label{div-v} 
\end{equation} 
and the surface conditions 
\begin{equation} 
\delta v^a \nabla_a h = 0, \qquad \eta^a \nabla_a h + \delta V + \delta U = 0 
\qquad (h=0). 
\label{surface_zerof} 
\end{equation} 
We recall that $\delta h = \delta V + \delta U$. 

The perturbed Euler equation (\ref{dEuler}) becomes 
\begin{equation} 
\nabla_a \delta V = -2\varepsilon_{abc} \Omega^b \delta v^c, 
\label{gradV_zerof} 
\end{equation} 
and taking the curl of both sides gives $\Omega^b \nabla_b \delta v^c = \Omega^a \nabla_b \delta v^b$. Accounting for Eq.~(\ref{div-v}), this is 
\begin{equation} 
\Omega^b \nabla_b \delta v^a = 0. 
\label{O_grad_dv} 
\end{equation} 
The perturbed Poisson equation (\ref{dPoisson}) gives rise to Eqs.~(\ref{mac3}) and (\ref{mac4}) --- unchanged --- and the surface condition of Eq.~(\ref{Dh}) is already captured by Eq.~(\ref{surface_zerof}). 

To proceed it is helpful to introduce the cylindrical coordinates $(s,\phi,z)$, which are defined in the usual way in terms of Cartesian coordinates, $x = s\cos\phi$, $y = s\sin\phi$, and $z = z$; $s$ is the distance from the rotation axis. In these coordinates we have that $\Omega^a = (0,0,\Omega)$ and $v^a = (0,\Omega,0)$. 

It follows from Eq.~(\ref{O_grad_dv}) that $\delta v^a$ must be independent of $z$. Because it must also be independent of $\phi$ (axisymmetry), we have that $\delta v^a$ is a function of only $s$. The first of Eqs.~(\ref{div-v}) produces $s^{-1} d(s v^s)/ds = 0$, from which it follows that $v^s = \mbox{constant}/s$; regularity on the axis requires $v^s = 0$. Next we make use of the first of Eqs.~(\ref{surface_zerof}), in which we insert Eq.~(\ref{enthalpy}) for $h$, suitably converted to cylindrical coordinates. The equation implies that $\delta v^z$ must be zero on the surface of the Maclaurin spheroid (except possibly at $z=0$). Because $\delta v^z$ is a function of only $s$, this necessarily implies that $\delta v^z$ vanishes everywhere inside the spheroid. We have obtained 
\begin{equation} 
\delta v^a = \bigl( 0, \delta\Omega(s), 0 \bigr). 
\end{equation} 
The axisymmetric, zero-frequency modes correspond to a differential rotation described by the shift $\delta\Omega(s)$ in angular velocity. 

With $\delta v^a$ determined, Eq.~(\ref{gradV_zerof}) allows us to find $\delta V$. The $z$-component of the equation tells us that $\delta V$ is independent of $z$, so that $\delta V = \delta V(s)$. The $s$-component reveals that
\begin{equation} 
2 \Omega\, \delta \Omega = \delta \Omega^2 = -\frac{1}{s} \frac{d \delta V}{ds}. 
\label{dO2} 
\end{equation} 
The perturbation of the gravitational potential is obtained by integrating Eqs.~(\ref{mac3}) and imposing the surface condition of Eq.~(\ref{mac4}). There is an infinity of solutions to these equations: the zero-frequency modes are highly degenerate. 

\subsection{Mode construction} 

To construct a basis of modes we begin with the gravitational potential, and proceed as in Sec.~\ref{sec:inertial}. We work in the $(\zeta,\mu,\phi)$ coordinates of Sec.~\ref{sec:coordinates}, and the solutions to Eq.~(\ref{mac3}) are given by Eqs.~(\ref{U_in}) and (\ref{U_out}). Equation (\ref{mac4}) then implies that $\delta V = \beta {\sf P}_\ell(\mu)$ on the surface of the Maclaurin spheroid, and we obtain  
\begin{equation} 
\frac{\zeta_0(1 - \zeta_0\,\arccot\zeta_0)}{P_\ell(i\zeta_0)\, iQ_\ell(i\zeta_0)}\, \alpha = \alpha + \beta 
\end{equation} 
after simplification. This condition determines the ratio $\beta/\alpha$. 

Equation (\ref{mac4}) restricts the surface value of $\delta V$ to be a Legendre polynomial. But $\delta V$ is a function of a single variable, $s$, the distance to the rotation axis. Recalling Eqs.~(\ref{Re_Rp}) and (\ref{coords_2}), we have that $s = R_{\rm e}(1 - \mu^2)^{1/2}$ on the surface. The global solution for $\delta V$ is therefore 
\begin{equation} 
\delta V = \beta {\sf P}_\ell(u), \qquad u = \pm \sqrt{1 - (s/R_{\rm e})^2}, 
\end{equation} 
with the upper sign applying in the upper hemisphere, and the lower sign in the lower hemisphere. From Eq.~(\ref{dO2}) we then get 
\begin{equation} 
\delta \Omega^2 = -\frac{\beta}{R_{\rm e}} \frac{{\sf P}_\ell'(u)}{u}, 
\label{dO2_sol} 
\end{equation} 
where a prime indicates differentiation with respect to the argument. Regularity at $u = 0$ ($s = R_{\rm e}$) requires $\ell$ to be even. Because the Legendre polynomial is then an even function, the choice of sign in $u$ becomes immaterial. Modes with odd values of $\ell$ are not defined.  

The modes with $\ell = 0$ and $\ell = 2$ are trivial. In the first instance ($\ell=0$) we have a shift in the focal radius $a$ (with $\zeta_0$ fixed), with 
\begin{equation} 
\alpha = 4\pi G\rho a\, \zeta_0 (1+\zeta_0^2)\, \arccot\zeta_0\, \delta a. 
\end{equation} 
This corresponds to a change in the mean radius $R$ and mass $M$ of the Maclaurin spheroid. In the second instance ($\ell=2$) we have a shift in $\zeta_0$ (with $a$ fixed), with 
\begin{equation} 
\alpha = \frac{2}{3} \pi G \rho a^2\, (1 + 3\zeta_0^2) \bigl[ (1+3\zeta_0^2)\arccot\zeta_0 - 3\zeta_0 \bigr]\, \delta \zeta_0. 
\end{equation} 
This corresponds to a change in $\Omega$, the spheroid's angular velocity. Nontrivial modes begin with $\ell = 4$. 

\subsection{Jordan chain and overlap integrals with the gravitoelectric force density}

The displacement vectors $\chi_0^a := \delta v^a$ and $\chi_1^a := \eta^a$ constitute a Jordan chain of length one associated with each axisymmetric, zero-frequency mode --- see Appendices A and D of Schenk {\it et al.}\ \cite{schenk-etal:01} for a description of Jordan chains. The zeroth member of the chain is a trivial displacement in the sense of Friedman and Schutz \cite{friedman-schutz:78a}: Because $\nabla_a(\rho \chi_0^a) = 0$, it gives rise to a vanishing density perturbation, and therefore to vanishing perturbations in the gravitational potential and specific enthalpy; and because $\partial_t \chi_0^a = 0$, it produces a vanishing perturbation in the velocity field. The first member of the chain, however, gives rise to a nonvanishing perturbation of the density. 

The axisymmetric, zero-frequency modes would have an impact on a tidally perturbed Maclaurin spheroid if $\chi_0^a$ and $\chi_1^a$ had nonvanishing overlap integrals with the tidal force densities. In Sec.~\ref{sec:driving} we saw that this could occur, in principle, in the case of the Newtonian force. We show here that the overlap integrals with $\chi_0^a$ always vanish, but that those with $\chi_1^a$ are nonzero when $\ell = 0$ and $\ell = 2$. Because these are the trivial modes encountered previously, describing unphysical shifts in the mass and angular velocity of the Maclaurin spheroid, the nonzero overlap integrals are of no concern. {\it The axisymmetric, zero-frequency modes have no impact on a tidally perturbed Maclaurin spheroid.} 

According to Eqs.~(\ref{EB_decomposition}) and (\ref{U}), the $m=0$ piece of the tidal potential is 
\begin{equation} 
\tilde{U}^{m=0}_{\rm tidal}  = \frac{1}{8} \sqrt{\frac{5}{\pi}}\, \tilde{\cal E}^0\, (x^2 + y^2 - 2z^2) 
= \frac{1}{8} \sqrt{\frac{5}{\pi}}\, \tilde{\cal E}^0\, a^2 \bigl[ 1 + \zeta^2 - (1 + 3\zeta^2) \mu^2 \bigr],  
\label{U_m0} 
\end{equation} 
and the corresponding force density is $\tilde{f}^{\rm el}_a = \nabla_a \tilde{U}$ --- we henceforth omit the decorating labels on $\tilde{U}$, to avoid cluttering the notation. Because $\tilde{U}$ is axisymmetric and the only nonvanishing component of $\chi_0^a = \delta v^a$ is along $\phi$, we have that 
\begin{equation} 
\tilde{f}^{\rm el}_a \chi^a_0 = 0, 
\end{equation} 
so that $\bkt{\chi_0}{f^{\rm el}} = 0$. As promised, all overlap integrals with $\chi_0^a$ vanish. 

The computation of $\bkt{\chi_1}{f^{\rm el}}$ might in principle require a detailed expression for $\chi_1^a = \eta^a$, which was not calculated. It is sufficient, however, to proceed on the basis of Eqs.~(\ref{div-v}) and (\ref{surface_zerof}). We exploit the fact that $\nabla_a \eta^a =0$ to integrate by parts, writing 
\begin{equation} 
\bkt{\chi_1}{f^{\rm el}} = \rho \int \eta^a \nabla_a \tilde{U}\, d{\cal V} 
= \rho \oint \tilde{U} \eta^a n_a\, dS, 
\end{equation} 
where $n_a$ is the outward unit normal and $dS$ the element of surface area. Writing $n_a = -|\bm{\nabla} h|^{-1} \nabla_a h$ and making use of Eq.~(\ref{surface_zerof}), we have that 
\begin{equation} 
\bkt{\chi_1}{f^{\rm el}} = \rho \oint |\bm{\nabla} h|^{-1} \tilde{U} (\delta V + \delta U)\, dS. 
\end{equation} 
We evaluate the surface integral by taking advantage of the $(\zeta,\mu,\phi)$ coordinates. The surface is situated at $\zeta = \zeta_0$, and in these coordinates $|\bm{\nabla} h| \propto (\zeta_0^2 + \mu^2)^{1/2}$ and $dS \propto (\zeta_0^2 + \mu^2)^{1/2}\, d\mu d\phi$. We also have that $\delta V + \delta U \propto {\sf P}_\ell(\mu)$, and  
Eq.~(\ref{U_m0}) gives $\tilde{U} \propto 2 - (1+3\zeta_0^2) {\sf P}_2(\mu)$. Integration yields 
\begin{equation} 
\bkt{\chi_1}{f^{\rm el}} \propto 2 \delta_{\ell,0} - \frac{1}{5} (1+3\zeta_0^2) \delta_{\ell,2}. 
\end{equation} 
As promised, the only nonvanishing overlap integrals are those with the trivial modes labeled $\ell = 0$ and $\ell = 2$.    

\section{Integrals of associated Legendre functions} 
\label{sec:legendre} 

The computations sketched in Sec.~\ref{sec:vanishing} require the evaluation of integrals over the members of the set $\cal S$ defined in Eq.~(\ref{set}). We list these integrals here, for $m=0$ and $m=2$. They can be established by differentiating the right-hand sides, and invoking the recursion relation of Eq.~(\ref{recursion1}), as well as [NIST Eq.~(14.10.5)] 
\begin{equation} 
(1-x^2) \frac{d {\sf P}_{\ell+1}^m}{dx} = -(\ell+1) x {\sf P}_{\ell+1}^m + (\ell+m+1) {\sf P}_\ell^m.  
\label{recursion2} 
\end{equation} 

For $m = 0$ we have 
\begin{subequations} 
\begin{align} 
\int {\sf P}_\ell\, dx &= -\frac{x}{\ell}\, {\sf P}_\ell + \frac{1}{\ell}\, {\sf P}_{\ell+1}, \\ 
\int x {\sf P}_\ell\, dx &= \biggl[ -\frac{x^2}{\ell-1} + \frac{1}{(\ell-1)(\ell+2)} \biggr] {\sf P}_\ell 
+ \frac{(\ell+1)x}{(\ell-1)(\ell+2)} {\sf P}_{\ell+1}, \\ 
\int x^2 {\sf P}_\ell\, dx &=\biggl[ -\frac{x^3}{\ell-2} + \frac{2(\ell+1)x}{(\ell-2)\ell(\ell+3)} \biggr] {\sf P}_\ell   
+ \biggl[ \frac{(\ell+1) x^2}{(\ell-2)(\ell+3)} - \frac{2}{(\ell-2)\ell(\ell+3)} \biggr] {\sf P}_{\ell+1}, \\ 
\int x^3 {\sf P}_\ell\, dx &= \biggl[ -\frac{x^4}{\ell-3} + \frac{3(\ell+1)x^2}{(\ell-3)(\ell-1)(\ell+4)} 
- \frac{6}{(\ell-3)(\ell-1)(\ell+2)(\ell+4)} \biggr] {\sf P}_\ell
\nonumber \\ & \quad \mbox{} 
+ \biggl[ \frac{(\ell+1) x^3}{(\ell-3)(\ell+4)} - \frac{6(\ell+1)x}{(\ell-3)(\ell-1)(\ell+2)(\ell+4)} \biggr] {\sf P}_{\ell+1} 
\end{align} 
\end{subequations} 
and 
\begin{subequations} 
\begin{align} 
\int x {\sf P}_{\ell+1}\, dx &= -\frac{(\ell+1)x}{\ell(\ell+3)}\, {\sf P}_\ell 
+ \biggl[ \frac{x^2}{\ell+3} + \frac{1}{\ell(\ell+3)} \biggr] {\sf P}_{\ell+1}, \\ 
\int x^2 {\sf P}_{\ell+1}\, dx &= \biggl[ -\frac{(\ell+1) x^2}{(\ell-1)(\ell+4)} 
+ \frac{2}{(\ell-1)(\ell+2)(\ell+4)} \biggr] {\sf P}_\ell 
+ \biggl[ \frac{x^3}{\ell+4} + \frac{2(\ell+1)x}{(\ell-1)(\ell+2)(\ell+4)} \biggr] {\sf P}_{\ell+1}. 
\end{align} 
\end{subequations} 

For $m = 2$ we have 
\begin{subequations} 
\begin{align} 
\int {\sf P}_\ell^2\, dx &= \frac{1}{1-x^2} \Biggl\{ 
\biggl[ \frac{x^3}{\ell} - \frac{(\ell^2+3\ell+6)x}{\ell(\ell+1)(\ell+2)} \biggr] {\sf P}_\ell^2 
+ \biggl[ -\frac{(\ell-1)x^2}{\ell(\ell+1)} + \frac{\ell^2+\ell+2}{\ell(\ell+1)(\ell+2)} \biggr] {\sf P}_{\ell+1}^2 \Biggr\}, \\ 
\int x {\sf P}_\ell^2\, dx &= \frac{1}{1-x^2} \Biggl\{ 
\biggl[ \frac{x^4}{\ell-1} - \frac{(\ell^2+4\ell+7)x^2}{(\ell-1)(\ell+1)(\ell+2)} 
+ \frac{\ell^2+\ell+4}{(\ell-1)\ell(\ell+1)(\ell+2)} \biggr] {\sf P}_\ell^2 
\nonumber \\ & \quad \mbox{} 
+ \biggl[ -\frac{x^3}{\ell+2} + \frac{(\ell^2+\ell+4)x}{\ell(\ell+1)(\ell+2)} \biggr] {\sf P}_{\ell+1}^2 \Biggr\}, \\ 
\int x^2 {\sf P}_\ell^2\, dx &= \frac{1}{1-x^2} \Biggl\{ 
\biggl[ \frac{x^5}{\ell-2} - \frac{(\ell+2)x^3}{(\ell-2)\ell} 
+ \frac{2(\ell^2+\ell+6)x}{(\ell-2)\ell(\ell+1)(\ell+2)} \biggr] {\sf P}_\ell^2 
\nonumber \\ & \quad \mbox{} 
+ \biggl[ -\frac{(\ell-1)x^4}{(\ell-2)(\ell+3)} + \frac{(\ell-1)(\ell^2+\ell+6)x^2}{(\ell-2)\ell(\ell+1)(\ell+3)} 
- \frac{2(\ell^2+\ell+6)}{(\ell-2)\ell(\ell+1)(\ell+2)(\ell+3)} \biggr] {\sf P}_{\ell+1}^2 \Biggr\}, \\ 
\int x^3 {\sf P}_\ell^2\, dx &= \frac{1}{1-x^2} \Biggl\{ 
\biggl[ \frac{x^6}{\ell-3} - \frac{(\ell^2+6\ell+3)x^4}{(\ell-3)(\ell-1)(\ell+4)} 
+ \frac{3(\ell+3)(\ell^2+\ell+8)x^2}{(\ell-3)(\ell-1)(\ell+1)(\ell+2)(\ell+4)} 
\nonumber \\ & \quad \mbox{} 
- \frac{6(\ell^2+\ell+8)}{(\ell-3)(\ell-1)\ell(\ell+1)(\ell+2)(\ell+4)} \biggr] {\sf P}_\ell^2 
+ \biggl[ -\frac{(\ell-1)x^5}{(\ell-3)(\ell+4)} + \frac{(\ell^2+\ell+8)x^3}{(\ell-3)(\ell+2)(\ell+4)} 
\nonumber \\ & \quad \mbox{} 
- \frac{6(\ell^2+\ell+8)x}{(\ell-3)\ell(\ell+1)(\ell+2)(\ell+4)} \biggr] {\sf P}_{\ell+1}^2 \Biggr\} 
\end{align} 
\end{subequations} 
and 
\begin{subequations} 
\begin{align} 
\int x{\sf P}_{\ell+1}^2\, dx &= \frac{1}{1-x^2} \Biggl\{ 
\biggl[ \frac{x^3}{\ell} - \frac{(\ell^2+3\ell+6)x}{\ell(\ell+1)(\ell+2)} \biggr] {\sf P}_\ell^2 
\nonumber \\ & \quad \mbox{} 
+ \biggl[ -\frac{x^4}{\ell+3} + \frac{(\ell^2+3)x^2}{\ell(\ell+1)(\ell+3)} 
+ \frac{\ell^2+3\ell+6}{\ell(\ell+1)(\ell+2)(\ell+3)} \biggr] {\sf P}_{\ell+1}^2 \Biggr\}, \\ 
\int x^2{\sf P}_{\ell+1}^2\, dx &= \frac{1}{1-x^2} \Biggl\{ 
\biggl[ \frac{(\ell+3)x^4}{(\ell-1)(\ell+4)} 
- \frac{(\ell+3)(\ell^2+3\ell+8)x^2}{(\ell-1)(\ell+1)(\ell+2)(\ell+4)} 
+ \frac{2(\ell^2+3\ell+8)}{(\ell-1)\ell(\ell+1)(\ell+2)(\ell+4)} \biggr] {\sf P}_\ell^2 
\nonumber \\ & \quad \mbox{} 
+ \biggl[ -\frac{x^5}{\ell+4} + \frac{\ell x^3}{(\ell+2)(\ell+4)} 
+ \frac{2(\ell^2+3\ell+8)x}{\ell(\ell+1)(\ell+2)(\ell+4)} \biggr] {\sf P}_{\ell+1}^2 \Biggr\}. 
\end{align} 
\end{subequations} 

\bibliography{/Users/poisson/writing/papers/tex/bib/master}
\end{document}